\documentclass[aps, prd, twocolumn, amsmath, floats, floatfix, superscriptaddress, nofootinbib]{revtex4-2}

\usepackage{graphicx}
\usepackage{epsfig}
\usepackage{amsmath,amsfonts,amssymb}
\usepackage{caption}
\usepackage{subcaption}
\usepackage{wrapfig}
\usepackage{verbatim}
\usepackage{float}
\usepackage{morefloats}
\usepackage[dvipsnames]{xcolor}
\usepackage{booktabs}
\usepackage[normalem]{ulem}
\usepackage{multirow}
\usepackage{makecell}
\usepackage{tabulary}
\usepackage{cancel}
\usepackage{xcolor}
\usepackage{blindtext}      
\usepackage{hyperref}       
\hypersetup{
	colorlinks   = true, 
	urlcolor     = blue, 
	linkcolor    = blue, 
	citecolor    = red,  
}
\usepackage{url}

\usepackage{booktabs}       
\usepackage{multirow}       
\usepackage{adjustbox}      
\usepackage{cellspace}
\usepackage{tablefootnote}  
\usepackage{physics}

\usepackage{enumitem}
\setlist[itemize]{noitemsep, topsep=0cm, leftmargin = 1.2cm}        
\setlist[enumerate]{noitemsep, topsep=0cm, leftmargin = 1.2cm}

\usepackage{ragged2e} 
\usepackage{subcaption} 
\DeclareCaptionJustification{justified}{\justifying} 

\begin{document}

\title{Rapidly rotating neutron star collapse in massive scalar-tensor theories}

\author{José Carlos Olvera M.}
\email{jose.olvera-meneses@uni-tuebingen.de}
\affiliation{Theoretical Astrophysics, IAAT, Eberhard Karls University of Tübingen, 72076 Tübingen, Germany}
\affiliation{Departamento de  Astronom\'{\i}a y Astrof\'{\i}sica, Universitat de Val\`encia,  Avinguda Vicent Andrés Estellés 19, 46100, Burjassot (Val\`encia), Spain}

\author{Daniela D. Doneva}
\affiliation{Theoretical Astrophysics, IAAT, Eberhard Karls University of Tübingen, 72076 Tübingen, Germany}
\affiliation{Departamento de  Astronom\'{\i}a y Astrof\'{\i}sica, Universitat de Val\`encia,  Avinguda Vicent Andrés Estellés 19, 46100, Burjassot (Val\`encia), Spain}

\author{Pablo Cerdá-Durán}
\affiliation{Departamento de  Astronom\'{\i}a y Astrof\'{\i}sica, Universitat de Val\`encia,  Avinguda Vicent Andrés Estellés 19, 46100, Burjassot (Val\`encia), Spain}
\affiliation{Observatorio Astron\'omico, Universitat de València, C/ Catedrático Jos\'e Beltr\'an 2, 46980, Paterna (València), Spain}

\author{José A. Font}
\affiliation{Departamento de  Astronom\'{\i}a y Astrof\'{\i}sica, Universitat de Val\`encia,  Avinguda Vicent Andrés Estellés 19, 46100, Burjassot (Val\`encia), Spain}
\affiliation{Observatorio Astron\'omico, Universitat de València, C/ Catedrático Jos\'e Beltr\'an 2, 46980, Paterna (València), Spain}

\author{Stoytcho S. Yazadjiev}
\affiliation{Department of Theoretical Physics, Faculty of Physics, Sofia University, Sofia 1164, Bulgaria}
\affiliation{Institute of Mathematics and Informatics, 	Bulgarian Academy of Sciences, 	Acad. G. Bonchev St. 8, Sofia 1113, Bulgaria}

\begin{abstract}
We present a full 3D numerical evolution code to study neutron stars in massive-scalar-tensor theories. The code is embedded in the \textsc{Einstein Toolkit} framework and its implementation constitutes a modified version of the Baumgarte-Shapiro-Shibata-Nakamura formalism with an additional nonminimally coupled scalar field. The approach we follow preserves the standard hydrodynamic evolution for matter fields, allowing eventually for a straightforward inclusion of more microphysical effects and better flexibility.  Using this code, we examine the gravitational collapse of rapidly rotating, scalarized neutron stars to a black hole by exploring the influence of the scalar field on the dynamical features of the process and on the gravitational-wave emission. We find that for the configurations studied in this work, there is an observational degeneracy in the tensorial gravitational-wave emission between collapsing scalarized stars and their counterparts in general relativity. However, this degeneracy can be broken through the emission of scalar radiation, which carries an energy of $\sim10^{-3}M_\odot c^2$. This is orders of magnitude higher than the quadrupolar emission ($\sim10^{-7}M_\odot c^2$) and might be used as an observational probe of modified gravity. We also find that rapid rotation can enhance this signal, since fast rotating stars can sustain larger scalar field amplitudes. 
\end{abstract}

\maketitle

%%%%%%%%%%%%%%%%%%%%%%
\section{Introduction}
%%%%%%%%%%%%%%%%%%%%%%

The investigation of the strong-field regime of gravity is becoming increasingly more accessible thanks to the efforts of several international collaborations. On one hand, very long baseline-interferometry observations conducted by the Event Horizon Telescope Collaboration have explored the environment surrounding supermassive black holes in nearby galaxies \cite{EventHorizonTelescope:2019dse,EventHorizonTelescope:2022wkp}. On the other hand, the LIGO-Virgo-KAGRA  (LVK) collaboration has already reported 218 confident gravitational-wave (GW) observations from coalescing compact binaries since the seminal first observation  (GW150914) and up to the first part of the fourth observing run O4a~\cite{TheLIGOScientificCollaboration2025, GWTC-4-update, LIGOScientific20254O}.  Ongoing efforts and updates to the current LVK detector network will allow in the upcoming years to cover the wide frequency range of the GWs emitted by different astrophysical sources, spanning from mHz to kHz. The high-frequency band ($\sim$ kHz) will be covered by the LVK Collaboration, Cosmic Explorer \cite{CosmicExplorer2022} and the Einstein Telescope \cite{ET_D_sensitivity}, while space-based missions such as DECIGO \cite{DECIGO2017}, TianQin\cite{TianQin} and LISA \cite{LISA:2022yao,LISA:2022kgy} will cover the low frequency band, the former observing in the $0.1-10$Hz range and the latter two missions in the $10^{-4}-1$Hz window. Moreover, efforts are also being made to conduct GW searches at MHz to GHz frequencies associated with early-universe physics or exotic high-frequency processes (see~\cite{Aggarwal:2025} and references therein). 
This expanded observational landscape will enable new tests of deviations from general relativity (GR) in the strong-field regime, with increasing accuracy as the sensitivity of the detectors improves. At present, the most recent tests of GR reported by the LVK Collaboration using the 91 confident binary black hole signals included in the fourth GW Transient Catalog (GWTC-4.0) show no evidence for deviations from GR~\cite{TGR1,TGR2,TGR3}. Nevertheless, the improved sensitivity of the next generation of gravitational wave detectors has the potential to open doors towards new tests of the strong field regime of gravity \cite{ET:2025xjr,LISA:2022kgy}.

Although GR is considered the standard framework for describing gravity, there is both theoretical and experimental evidence suggesting that this theory is not fully complete and must be extended or generalized \cite{Berti_2015}. This is especially true when dealing with strong fields, such as in the interior of black holes, the early Universe, or the dynamical formation of singularities that occur during gravitational collapse \cite{Grav_collapse_penrose_singularitye}. In order to address these limitations, different flavors of modifications and extensions to GR have been proposed, offering new ways to investigate these extreme regimes and the physics of black holes, dark matter, and dark energy \cite{galaxies10040076}. 

One of the most prominent generalizations of Einstein's theory of gravity is the family of scalar-tensor theories (STT) \cite{DEF_multiscalar1992}. Such theories involve a scalar field which can be viewed as a mediator of gravitational interaction in addition to the spacetime metric \cite{actionBD,actionBD1,DEF_multiscalar1992}. In the simplest non-trivial case where this scalar field is massless and only carries a kinetic term, STTs have already been well constrained through Solar System and binary pulsar experiments \cite{Will:2014kxa,Zhao:2022vig,Freire:2024adf}. However, through the inclusion of  a mass term, a new layer of freedom is added to study possible alternatives and pass observational tests~\cite{Ramazanolu2016,Yazadjiev:2016pcb}. 

A good starting point to test and constrain beyond-GR theories is to study the effects of such modifications on the dynamical behavior and GW emission of compact objects. One family of such objects is neutron stars. Despite the presence of matter at supra-nuclear density and the related equation-of-state (EoS) uncertainties, neutron stars offer a broader variety of observable astrophysical phenomena than black holes, making them in some cases the best probes of strong gravity  \cite{Zhao:2022vig,Silva:2020acr,LIGOScientific:2017zic}. Neutron stars can also act as a formation channel for black holes through gravitational collapse. Simulating this highly dynamical process was one of the most challenging problems at the dawn of numerical relativity, owing to both computational and methodological limitations (see~\cite{Nakamura:1981,Stark-Piran:1985}). Over the past few decades, this process has been extensively studied within GR and is now regarded as a standard benchmark for testing new numerical relativity codes. Analytic studies were first reported in \cite{Collapse_perturbations} while numerical approaches have been presented both in spherical symmetry \cite{CollapseNovakGR} as in full 3D  \cite{collapse_1st,3dcollapse1,CollapseReisswig2012ThreeDimensionalGH,Collapse_Dietrich_Bernuzzi}.

Of particular interest is the collapse of rapidly rotating neutron stars, as it enables studies of specific  astrophysical processes such as gravitational-wave emission \cite{3dcollapse2GW} and gamma-ray bursts \cite{disk_collapse,collapse_disk1}. 

In the context of STT, studies of gravitational stellar collapse remain restricted to the spherically symmetric case \cite{STTcollpse,corecollapse_scalar}.  Early work focused on analytic approaches to compute perturbations associated with the scalar field \cite{STTcoll_perturbations}, followed by 1D simulations of Oppenheimer-Snyder dust collapse \cite{BDcollapse_dust,BDcollapse_radiation}. The first simulation considering a neutron star modeled through a polytropic EoS was presented in \cite{STTcollpse}, assuming a massless field and spherical symmetry.  
Subsequent studies extended the framework of~\cite{STTcollpse} to massive scalar fields, investigating the transition from a stellar core (a proto-neutron star) to a black hole under spherical symmetry through core-collapse simulations \cite{Sperhake:2017itk,corecollapse_scalar} assuming different types of couplings or interaction potentials for the scalar sector \cite{Rosca-Mead:2019seq,Geng:2020slq,Asakawa:2023obq,scalarcollapse}. Spherically symmetric stellar collapse in other modified theories of gravity was considered as well \cite{Kuan:2021lol}.  Recently, the first core collapse supernova simulation (CCSN) in massive scalar tensor theories (MSTT) within axisymmetry incorporating neutrino treatment was performed \cite{Kuroda:2023zbz}. Using a spherically symmetric progenitor model, their work yielded a spherical explosion and demonstrated how scalar field dynamics leave imprints on the GW emission, neutrino signals, and the explosion dynamics. 

In addition to these dynamical studies, such core-collapse events in MSTT can act as a strong, long-lived source of gravitational waves with scalar polarizations, resulting in a detectable stochastic background, even in spherical symmetry \cite{Rosca-Mead:2023tdc}.   However, to date, and to the best of our knowledge, the collapse of a rotating progenitor in MSTT remains unexplored, leaving important open questions about the role of multidimensionality and the scalar effects in core collapse.

As a first step toward addressing this gap, we investigate the role of rotation in one possible outcome of CCSN: the collapse of a rotating, scalarized neutron star into a black hole.  To this end, we have developed an extension of the \textsc{Einstein Toolkit} capable of simulating such collapses within these theories.
Our new code simulates the gravitational dynamics of matter and spacetime in the presence of a massive scalar field coupled to the geometric sector. The implementation is based on a modified version of the Baumgarte–Shapiro–Shibata–Nakamura (BSSN) formalism \cite{BSSN_BS,BSSN_SN}, extended to incorporate a scalar–spacetime coupling following \cite{Shibata-2014}.
We simulate uniformly rotating, scalarized neutron stars and compare their collapse with GR counterparts, focusing on the global dynamics of the collapse, black hole formation, and GW emission. In this first study, we adopt a simple polytropic EoS with a thermal hybrid component; more realistic matter models will be considered in future work. Despite this limitation, our results provide a necessary first step toward understanding rotating neutron star collapse in MSTTs.

This paper is organized as follows. Section 2 presents a brief introduction to scalar-tensor theories and to the mathematical and physical framework on which our implementation is based. The generation of initial data, as well as the description of the models used in this work, is described in Section 3. Section 4 details the numerical aspects of the code. The results of the simulations are discussed in Section 5 and the conclusions are reported in Section 6. The paper also contains four appendices where additional technical information on the mathematical framework and on the numerics is included.

%%%%%%%%%%%%%%%%%%%%%%%%%%%%%%%%%
\section{scalar-tensor theories}
%%%%%%%%%%%%%%%%%%%%%%%%%%%%%%%%%

\subsection{Jordan and Einstein frames} 
 In scalar-tensor theories, the gravitational interaction is mediated not only by the spacetime metric, denoted here by $g_{\mu\nu}$, but also by a dynamical scalar field $\Phi$, the so-called gravitational scalar. The most general action of the scalar-tensor theories in the physical Jordan frame is given by \cite{actionBD,actionBD1} 
\begin{equation}
\begin{aligned}
S_{\text{J}} = & \frac{1}{16\pi G_*}\int d^4x \sqrt{-g}
\Big[  \Phi R  \\
&  - \frac{\omega(\Phi)}{\Phi}
g^{\mu\nu} \nabla_\mu\Phi\nabla_\nu\Phi
- U(\Phi) \Big] \\
& +\, S_{\text{matter}}(\Psi_{\text{matter}}, g_{\mu\nu})\,.
\end{aligned}
\label{action}
\end{equation}
Here, $G_{*}$ is the bare gravitational constant, $\nabla_\mu$ is the covariant derivative, and $R$ is the Ricci scalar curvature with respect to the spacetime metric $g_{\mu\nu}$. The dynamics of the gravitational scalar $\Phi$ is governed by the functions $\omega(\Phi)$ and $U(\Phi)$. In order for the graviton to carry positive energy we must have $\Phi>0$. The inequality $2\omega(\Phi) + 3\ge 0$ ensures that the kinetic energy of the scalar field is non-negative. Moreover, $S_{\text{matter}}$ in Eq.~(\ref{action}) is the action of the matter fields which are collectively denoted by $\Psi_{\text{matter}}$. In what follows we shall use $G_{*}=1$. 

Varying the action with respect to the spacetime metric and the scalar field leads to the following Jordan frame field equations:
\begin{equation}\label{eq:EinsteinEqs}
    \begin{aligned}
       G_{\mu\nu} = \frac{8\pi}{\Phi}\left( T^{\rm matter}_{\mu\nu} + T^{\rm SF}_{\mu\nu}\right)\,,
    \end{aligned}
\end{equation}
\begin{equation}\label{eq:SF}
    \begin{aligned}
        \Box \Phi = \frac{1}{2\omega(\Phi)+3}\left[ 8\pi T - \frac{d\omega}{d\Phi}\nabla_\alpha\Phi\nabla^\alpha\Phi \right. \\ +  \Phi^3\frac{d}{d\Phi} \left.\left(\frac{U(\Phi)}{\Phi^2} \right)\right]\,,
    \end{aligned}
\end{equation}
where $\Box$ is the d'Alembertian operator and $T$ is the trace of the energy-momentum tensor, with
the effective energy-momentum tensor of the scalar field, $ T^{\rm SF}_{\mu\nu}$ given by 
\begin{equation}\label{eq:SF_T}
    \begin{aligned}
      8 \pi T^{\rm SF}_{\mu\nu} = \frac{\omega(\Phi)}{\Phi} &\left( \nabla_\mu \Phi \nabla_\nu \Phi  -\frac{1}{2} g_{\mu\nu} \nabla_\alpha\Phi\nabla^\alpha\Phi  \right) \\+ & \nabla_\mu\nabla_\nu \Phi -g_{\mu\nu} \Box \Phi-\frac{1}{2} g_{\mu\nu} U(\Phi)\, .
    \end{aligned}
\end{equation}
From a mathematical point of view it is also convenient to analyze scalar-tensor theories with respect to the so-called Einstein frame where the tensor and the scalar degrees of freedom are decoupled.  The  Einstein frame is defined by the metric
\begin{equation}
g^*_{\mu\nu}= \Phi g_{\mu\nu}\,.
\end{equation}
We also introduce the new scalar field $\phi$ via the equation 
\begin{equation}
\left(\frac{d\phi}{d\Phi}\right)^2 = \frac{3 + 2\omega(\Phi)}{4\Phi^2}\,,
\end{equation}
and define 
\begin{equation}
A(\phi)=\Phi^{-1/2} (\phi), \;\;\; 4V(\phi)= U(\Phi(\phi)) \Phi^{-2}(\phi),
\end{equation}
where $A(\phi)$ plays the role of a conformal factor relating the Einstein and the Jordan frame metrics. Through the above definitions, the Jordan frame action can be transformed into the Einstein frame,
\begin{equation}
\begin{aligned}    S_{\text{E}} =& \frac{1}{16\pi G}  \int d^4x \sqrt{-g^*} \left[ R^* - 2g^{*\mu\nu}\partial_\mu\phi\partial_\nu\phi\right. \\ & \left.- 4V(\phi)\right] + \, S_{\text{matter}}(\Psi_{\text{matter}}, A^2(\phi)g^*_{\mu\nu}).\\
\end{aligned}
\end{equation}

Varying the Einstein frame action with respect to the Einstein frame metric $g^*_{\mu\nu}$ and the scalar field $\varphi$ we obtain the following field equations:
\begin{equation}\label{eq:Einstein_frame}
    \begin{aligned}
       G^*_{\mu\nu} = & 8\pi T^{*}_{\mu\nu} + 2 \nabla^{*}_\mu\phi \nabla^{*}_\nu\phi \\ &- g^*_{\mu\nu} g^{\alpha\beta}_* \nabla^{*}_\alpha\phi\nabla^{*}_\beta\phi - 2V(\phi)g^{*}_{\mu\nu}\,,
 \end{aligned}
\end{equation}
\begin{equation}\label{eq:EinsteiF_sf}
    \begin{aligned}
      \Box_{*} \phi= - 4\pi \alpha(\phi)T^* + \frac{dV(\phi)}{d\phi} . 
 \end{aligned}
\end{equation}
Here, the superscript or subscript $*$ is used to indicate that the corresponding quantity is in the Einstein frame. $T^*$  is the trace of the (Einstein frame) energy momentum tensor and the function $\tilde{\alpha}(\varphi)$ is the logarithmic derivative of the conformal factor $A(\varphi)$,
\begin{equation}\label{eq:EFcoupling}
\tilde{\alpha}(\phi)= \frac{d\ln A(\phi)}{d\phi}.
\end{equation}
The Einstein frame energy-momentum tensor $T^*_{\mu\nu}$ is related to the Jordan frame one $T_{\mu\nu}$ via the relation $T^*_{\mu\nu}=A^2(\phi)T_{\mu\nu}$. 

In the weak-field limit, when the asymptotic cosmological value of the scalar field is $\phi_0$, the parameters $\alpha_0=\tilde{\alpha}(\phi_0)$ and 
$\beta_0=\frac{d\tilde{\alpha}}{d\phi}(\phi_0)$   are related directly with the parametrized post-Newtonian coefficients \cite{DEF1}, allowing to constrain these parameters through weak-field \cite{Will:2014kxa} and binary pulsar observations \cite{binarypulsar_constraints}. Although in our approach (and code) $\phi_0$ is a freely specifiable parameter, we will consider $\phi_0=0$ throughout this work for simplicity, since $\phi_0$ is strongly constrained by observations and typically does not lead to qualitatively new results.

In the context of neutron stars, the Einstein frame representation is particularly useful when considering deviations from GR, as the scalar field acts only as an additional source for the standard Einstein tensor $G_{\mu\nu}$ in the Einstein equations. On the other hand, the Jordan frame representation is convenient when dealing with matter, as it follows the same geodesics as in GR, in contrast to the Einstein frame, where matter is subjected to an additional scalar interaction.
In our work, we have implemented the evolution equations in the Jordan frame primarily for technical reasons. However, all physical predictions should be frame-independent.

%%%%%%%%%%%%%%%%%%%%%%%%%%%%%%%%%%%%%%%%%%%%%%%%%%%%%%%%%
\subsection{Specifying the scalar-tensor theory }
%%%%%%%%%%%%%%%%%%%%%%%%%%%%%%%%%%%%%%%%%%%%%%%%%%%%%%%%%

The scalar-tensor theory is fully specified by the functions $A(\phi)$ and $V(\phi)$. As we already mentioned, we shall consider here MSTTs with a mass term in the potential. For this reason, we chose  $V(\phi)$ as,
\begin{equation}\label{eq:EF_potential}
V(\phi)= \frac{1}{2}m^2_{\phi}\phi^2\,,
\end{equation}
with $m_\phi$ being the mass of the scalar field. For the conformal factor $A(\phi)$ we consider the expression 
\begin{equation}\label{eq:Couplingfunc}
A(\phi)= e^{\alpha_0\phi + \frac{1}{2}\beta_0\phi^2},  
\end{equation}
where $\alpha_0\ge 0$ and $\beta_0\le 0$ are parameters.  We consider $\beta_0\le 0$ because this is the standard case allowing for spontaneous scalarization \cite{Review_scalarization}, although it has also been shown that  $\beta_0> 0$ can also lead to scalarized solutions  for realistic nuclear matter at high enough densities \cite{Shao_scalar_core}.
We can restrict ourselves to  $\alpha_0\ge 0$ without loss of  generality since the scalar-tensor theory under consideration is invariant under the transformation
$(\alpha_0,\phi)\to (-\alpha_0,-\phi)$ (see \cite{Staykov:2026ojk} for a thorough exploration of the solution branches for the coupling \eqref{eq:Couplingfunc}).

In the particular case when $\beta_0=0$ the theory reduces to the massive Brans-Dicke (BD) theory while for $\alpha_0=0$ we recover the massive  Damour-Esposito-Farese (DEF) scalar-tensor theory, which allows for spontaneous scalarization and is indistinguishable  from GR
in the weak field limit. 

In terms of Jordan frame our theory corresponds to
\begin{eqnarray}
&&\frac{1}{3 + 2\omega(\Phi)}= \alpha^2_0 - \beta_0 \ln\Phi, \label{eq:JFcoupling} \\
&&U(\Phi)= \frac{2m^2_\Phi}{\beta^2_0}\left[\sqrt{\alpha^2_0 - \beta_0\ln\Phi} -\alpha_0\right]^2\Phi^2,   
\end{eqnarray}
where $m_\Phi=m_\phi$, since the physical mass of the scalar field is frame-invariant as one can easily check. 

Another consequence of having a massive field, is that the scalar field now has a finite range of interaction. In contrast with the massless case, where the field decays as $\sim1/r$ having an effectively infinite range of interaction, a massive field decays exponentially in a scale given by $\lambda_\Phi=\hbar c/m_\Phi$. This characteristic length scale $\lambda_\Phi$ represents the Compton wavelength associated to the fundamental frequency of the scalar field. This implies that for distances much larger than $\lambda_\Phi$, the scalar field is effectively suppressed.

Let us very briefly discuss the observational constraints on the parameters of our scalar-tensor theories. In the pure massless case $m_\Phi=0$, the theory is severely constrained by pulsar timing experiments in neutron star-white dwarf (NS-WD) binaries and NS–NS binaries, and there remains a very small window for the parameters \cite{Zhao:2022vig,Freire:2024adf}. The inclusion of a nonzero scalar mass $m_\Phi$ changes the picture considerably and for $m_\Phi \ge 10^{-16} eV$ pulsar timing can not in practice impose constraints \cite{Ramazanolu2016,Yazadjiev:2016pcb}. The presence of a scalar field will also leave imprints on the GW emission from binary neutron star mergers (e.g. GW170817). For roughly $m_\Phi\ge 10^{-12} {\rm \,eV}$, though, the current merger events can not strongly constrain the $(\alpha_0,\beta_0)$ parameter space \cite{Kuan:2023hrh}.  In this work we will limit ourselves to study the case of $m_\Phi=1.33\times10^{-11}$ \text{  eV/c$^2$}, which results in a Compton wavelength of around $15$ km.

We will follow the methodology from \cite{Shibata-2014, MST_postmerger_remnants}, where the numerical calculations are performed on the physical Jordan frame but in terms of the  rescaled field $\varphi=\sqrt{-2\beta_0}\phi$. We also define the  parameter $\delta_0= -\alpha_{0}\sqrt{-\frac{2}{\beta_0}}$ such that the gravitational scalar $\Phi$  is then given by: \begin{equation}
\Phi=e^{\delta_0\varphi + \frac{1}{2}\varphi^2}.  
\end{equation}

%%%%%%%%%%%%%%%%%%%%%%%%%%%%%%%%%
\subsection{Scalarization in neutron stars}
%%%%%%%%%%%%%%%%%%%%%%%%%%%%%%%%%
As commented above, our coupling function \eqref{eq:Couplingfunc} comprises two particular cases -- the BD and the DEF theory. In the former case, compact objects are always endowed with a scalar field and that is why even weak field observations can put strong constraints on the theory \cite{Will:2014kxa}. The DEF theory, though, offers a qualitatively new mechanism for developing a nontrivial scalar field where the weak field regime completely coincides with GR, the so-called spontaneous scalarization \cite{DEF1}.

The essence of spontaneous scalarization is that when the compactness of a neutron star exceeds a critical threshold, the GR solutions (solutions with a trivial  scalar field) become unstable. This solution will then migrate dynamically to a configuration of a neutron star with a nontrivial scalar field.  This phenomenon resembles a second-order phase transition.

This effect was first described in \cite{DEF1} for the massless case with a coupling function $A(\phi)=e^{\frac{1}{2}\beta_0\phi^2}$. In this framework, linear perturbations of the scalar field $\phi$ around a GR background satisfy an effective Klein–Gordon equation (Eq.~(\ref{eq:EinsteiF_sf})) with an effective mass given by $\mu_\mathrm{eff}^2=-4\pi G\beta_0 T$, where $T$ represents the trace of the energy-momentum tensor. The sign of $\mu_{\mathrm{eff}}^{2}$ is controlled by the coupling parameter $\beta_{0}$  and the trace of the energy momentum tensor $T$. Thus $\mu_{\mathrm{eff}}^{2}$ depends on the theory coupling parameters, the central density, and the EoS.  When $\mu_{\mathrm{eff}}^{2}<0$, the scalar field experiences a tachyonic instability that drives exponential growth of scalar perturbations until the non‑linear terms in the field equations become important and saturate the growth. The end result is a ``scalarized" configuration in which the neutron star develops a stable scalar hair. This spontaneous scalarization produces a new family of stars with a stellar structure that is different from the purely GR case \cite{Review_scalarization}.

Although it is possible to obtain scalarized solutions independently of the sign of $\beta_0$ \cite{Shao_scalar_core}, in this work we consider a negative $\beta_0$, for which the scalarization threshold appears when $T<0$. This means that for different nuclear matter models there are different values of $\beta_{0}$ where the effect is active. For stellar configurations where the tachyonic instability is not triggered, the scalar field remains constant across the star, and the stellar structure of GR is recovered. This is not the case when considering $\alpha_0\neq0$, as even if $\mu_{\mathrm{eff}}^{2}>0$, the stellar structure will be under the influence of a non-trivial scalar field configuration. In order to recover a weak-limit consistent with GR for $\alpha_0\neq0$, the value of $\alpha_0$ must be small enough so that the stellar models are indistinguishable with respect to GR \cite{scalarization2,Will:2014kxa}. 

In the case of the combined presence of both $\beta_0$ and $\alpha_0$ terms, spontaneous scalarization is strictly speaking not possible since GR is not a solution of the field equations and neutron stars will always be endowed with a scalar field.  For small enough $\alpha_0$, though, as the one considered in the present paper, we can have a picture that resembles spontaneous scalarization closely. Namely, at some critical central energy density, the star will jump from a weakly scalarized to a strongly scalarized state, thus developing strong, potentially detectable deviations from GR \cite{Ramazanolu2025}.

The inclusion of a mass term from the potential in Eq.~(\ref{eq:EF_potential}) results in the effective mass for scalarization being modified as $\mu_\mathrm{eff}^2=m_\Phi^2-4\pi G\beta_0 T$. Now, in order to obtain scalarization, the condition that must be fulfilled is $4\pi G\beta_0 T>m_\Phi^2$. As a consequence, higher absolute values of $\beta_0$ are required in order to trigger the tachyonic instability.

%%%%%%%%%%%%%%%%%%%%%%%%%%%%%%%%%
\subsection{MSTTs in the BSSN formalism}
%%%%%%%%%%%%%%%%%%%%%%%%%%%%%%%%%

We consider a spacetime foliation consisting of a family of spacelike hypersurfaces normal to a timelike unit vector $n^\mu$. This foliation is characterized by the lapse function $\alpha$ and the shift vector $\beta^i$. Assuming this mixed DEF+BD theory from  Eq.~(\ref{eq:Couplingfunc}), a $3+1$ decomposition of Eq.~(\ref{eq:SF}) along with the definition of the conjugate momentum associated with the scalar field, $\Pi_\Phi = -\mathcal{L}_n\Phi= - n^\mu \nabla_\mu\Phi$, results in a set of equations that can be expressed in terms of the scalar variable $\varphi$, 
 \begin{equation}\label{eq:SF3+1}
   \begin{aligned}
   (\partial_t-\beta^i\partial_i)\varphi = & -\alpha K^\varphi\,,
   \\
            (\partial_t-\beta^i\partial_i)K^\varphi = & -\alpha D_i\partial^i\varphi +\alpha KK^\varphi - \partial_i\alpha\partial^i\varphi  \\
            &-\alpha(\varphi+\delta_0)(\partial_i\varphi\partial^i\varphi-(K^\varphi)  ^2) \\&- 4\pi\alpha \beta_0 T (\varphi+\delta_0) e^{-\delta_0\varphi-\varphi^2/2} \\ &+ \alpha m_\Phi^2 \varphi e^{\delta_0\varphi+\varphi^2/2} \, ,
    \end{aligned}
\end{equation}
where $K^\varphi=-\mathcal{L}_n\varphi$ acts as an equivalent to $\Pi_\Phi$ for the variable $\varphi$. This transformation for the scalar field is particularly useful for recovering a Klein-Gordon equation at large distances, as discussed in~\cite{Shibata-2014}.  The rest of the evolution equations are shown in Appendix A. 

The BSSN formulation of scalar-tensor theories in the Jordan frame has been shown to be stable and suitable for performing numerical simulations of neutron stars, both for isolated and binary systems \cite{Shibata-2014}.  By implementing this approach within the \textsc{Einstein Toolkit} infrastructure \cite{EinsteinToolkit:2025_05}, the code developed in this work provides a common numerical framework for studying scalar–tensor theories while minimizing the need to modify the existing infrastructure. This also enables compatibility with the initial data routines, hydrodynamic and MHD modules, and analysis codes already available as part of the \textsc{Einstein Toolkit}. We plan to make this implementation publicly available as a set of computational modules, or ``thorns," for the \textsc{Einstein Toolkit}.

%%%%%%%%%%%%%%%%%%%%%%%%%%%%%%%%%%%%%%%%%%%%%%%%%%%%%%%%%%%%%%%%
\section{Initial data for rotating neutron stars in STT}
%%%%%%%%%%%%%%%%%%%%%%%%%%%%%%%%%%%%%%%%%%%%%%%%%%%%%%%%%%%%%%%%

In order to generate initial data for rotating neutron stars in STT we employ a modified version of the RNS code \cite{RNS_Stergioulas}, which is based on the Komatsu-Eriguchi-Hachisu (KEH) method \cite{KEH1,KEH2} with the modifications introduced in \cite{Cook:1992}. This allows us to obtain a solution for neutron stars in hydrostatic equilibrium within the specific MSTT parametrization discussed above \cite{rapidlyrotNSEJ,Doneva:2016xmf}. 

Our modified version of the RNS code is able to generate equilibrium solutions in MSTT by considering a stationary, axisymmetric, circular spacetime, i.e., a metric of the form 
\begin{equation}{\label{eq:RNS_metric}}
\begin{aligned}
        ds^2 = -e^{\xi+\sigma} dt^2 + e^{2\Delta} (dr^2 + r^2 d\theta^2) \\+ e^{\xi-\sigma}r^2 \sin^2\theta (d\phi - \omega dt)^2,
\end{aligned}
            \end{equation}
where the quantities $\xi$, $\sigma$, $\omega$ and $\Delta$ represent the metric potentials.  All quantities associated with the geometry of the spacetime are calculated in the Einstein frame, while the variables related to the matter content are given in the Jordan frame.

The RNS code simultaneously solves the Einstein equations, the scalar field equation, and the hydrostationary equilibrium equation for a given EoS. As a result, one obtains the metric potentials from Eq.~(\ref{eq:RNS_metric}), the distribution of the scalar field and the hydrodynamic variables.  The output of the RNS computation is a 2D stationary configuration which is then interpolated into a 3D Cartesian grid to serve as initial data for our simulations. These quantities are then transformed into the required 3+1 formalism variables, namely the metric and the extrinsic curvature. Details on how those are calculated in the Cartesian 3+1 decomposition are available in Appendix C. 

\begin{figure}[t]
    \centering
\includegraphics[width=0.99\linewidth]{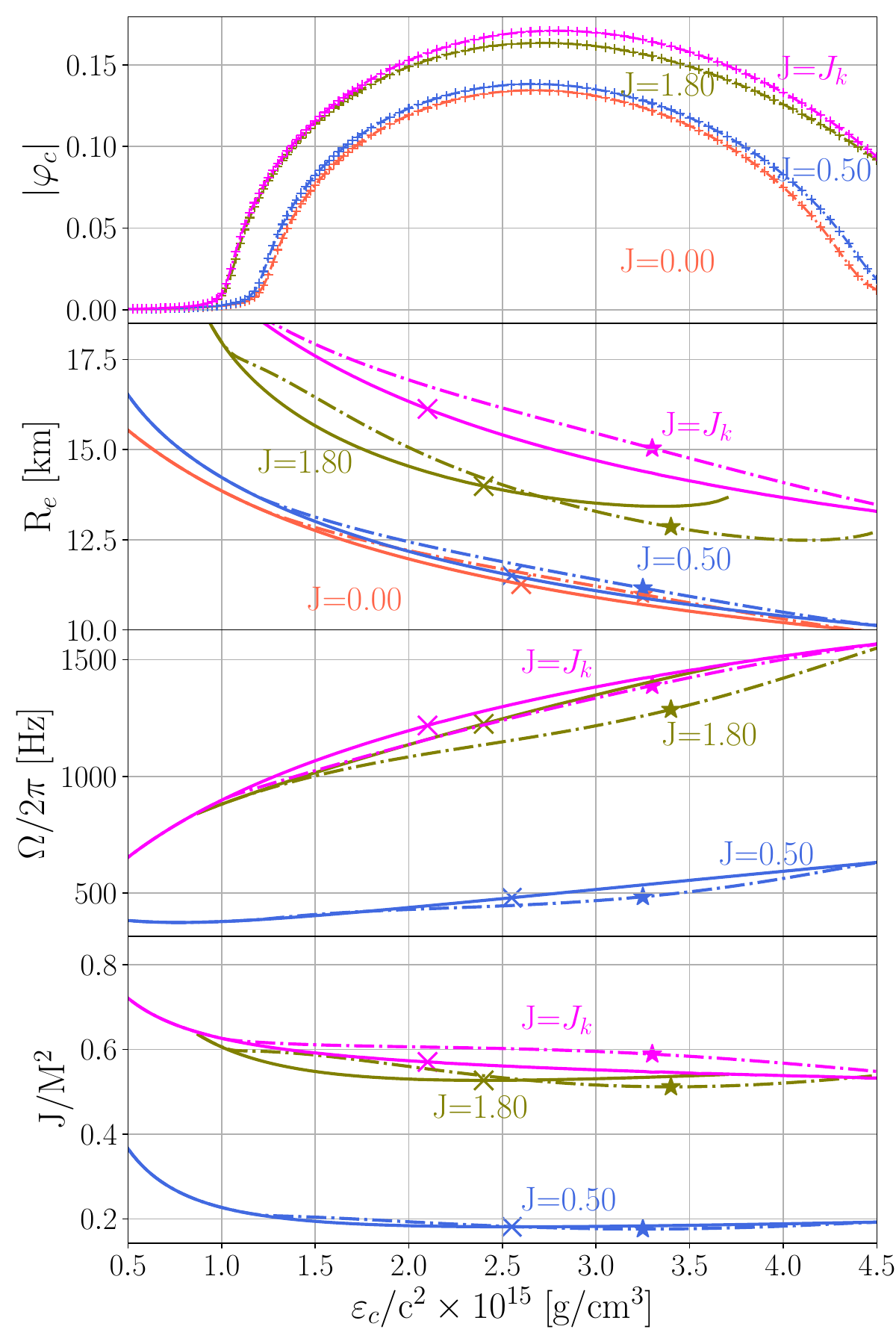}  
\caption{ Constant angular momentum sequences for rotating polytropic stars in MSTT, with $\alpha_0=10^{-3}$, $\beta_0=-7.5$ and $m_\Phi=1.33\times10^{-11}$ eV/c$^2$. This figure displays the degree of scalarization through the magnitude of the central scalar field (top panel),  the circumferential equatorial radius of the star (second row panel),  the rotational frequency (third row panel) and the spin parameter $J/M^2$ (bottom panel), where $J$ is the angular momentum and $M$ the mass of the star. In all panels solid lines represent the GR solution while dash-dotted lines show the scalarized solution. Stars and crosses represent the maximum mass solution for MSTT and GR, respectively. }    
\label{fig:Mass-radius}
\end{figure}\vspace{.1cm}

In GR, each stellar model can be uniquely specified through two input quantities: the central energy of the star $\varepsilon_c$ and the ratio between the polar and equatorial radius $r_p/r_e$. The latter describes the oblateness of the star and is a measure of how fast the star is rotating. In MSTT, it is necessary to consider three additional parameters: the mass of the scalar field $m_\Phi$ and the two coupling parameters $\alpha_0$ and $\beta_0$. This means that we require a combination of 5 parameters in order to fully specify a stellar model once the EoS is fixed. Throughout this work we follow the approach of previous studies on neutron star collapse in GR \cite{3dcollapse1,Collapse_Dietrich_Bernuzzi} where a polytropic EoS is adopted. We consider a polytropic constant $K=100$ and index $N=1$. This choice serves two purposes. First, it allows for a validation of our results against existing literature in both GR and STT \cite{Collapse_Dietrich_Bernuzzi,STTcollpse}, and second, it allows to isolate the effects from MSTT without introducing additional uncertainties due to nuclear matter. An extension of this work employing realistic (microphysical) EoS will be presented in future work.     

Since we want to be able to recover the weak limit of GR, we choose configurations with a fixed value of $\alpha_0= 10^{-3}$, for which only weakly scalarized solutions can exist $[\varphi \sim \mathcal{O}(10^{-3})]$ for the case $\beta_0=0$. For such small values of the scalar field, these solutions are indistinguishable from their GR counterparts for most practical purposes. Namely, we find that the stellar properties of the star do not deviate more than $0.002\%$ from the GR case. This value of $\alpha_0$ is also consistent with the latest upper bound of $\alpha_0<6\times10^{-3}$ imposed by binary pulsars in the case when $m_\Phi=0$ \cite{Zhao:2022vig}. For values of $\beta_0$ that are negative enough to induce scalarization, though, we can obtain strongly scalarized solutions $[\varphi\sim \mathcal{O}(10^{-1})]$, where the stellar structure can be altered dramatically with respect to the equivalent solution in GR. In this work, we focus only on stars that are already on the strongly scalarized branch.

\begin{figure}[t]
    \centering
\includegraphics[width=0.9\linewidth]{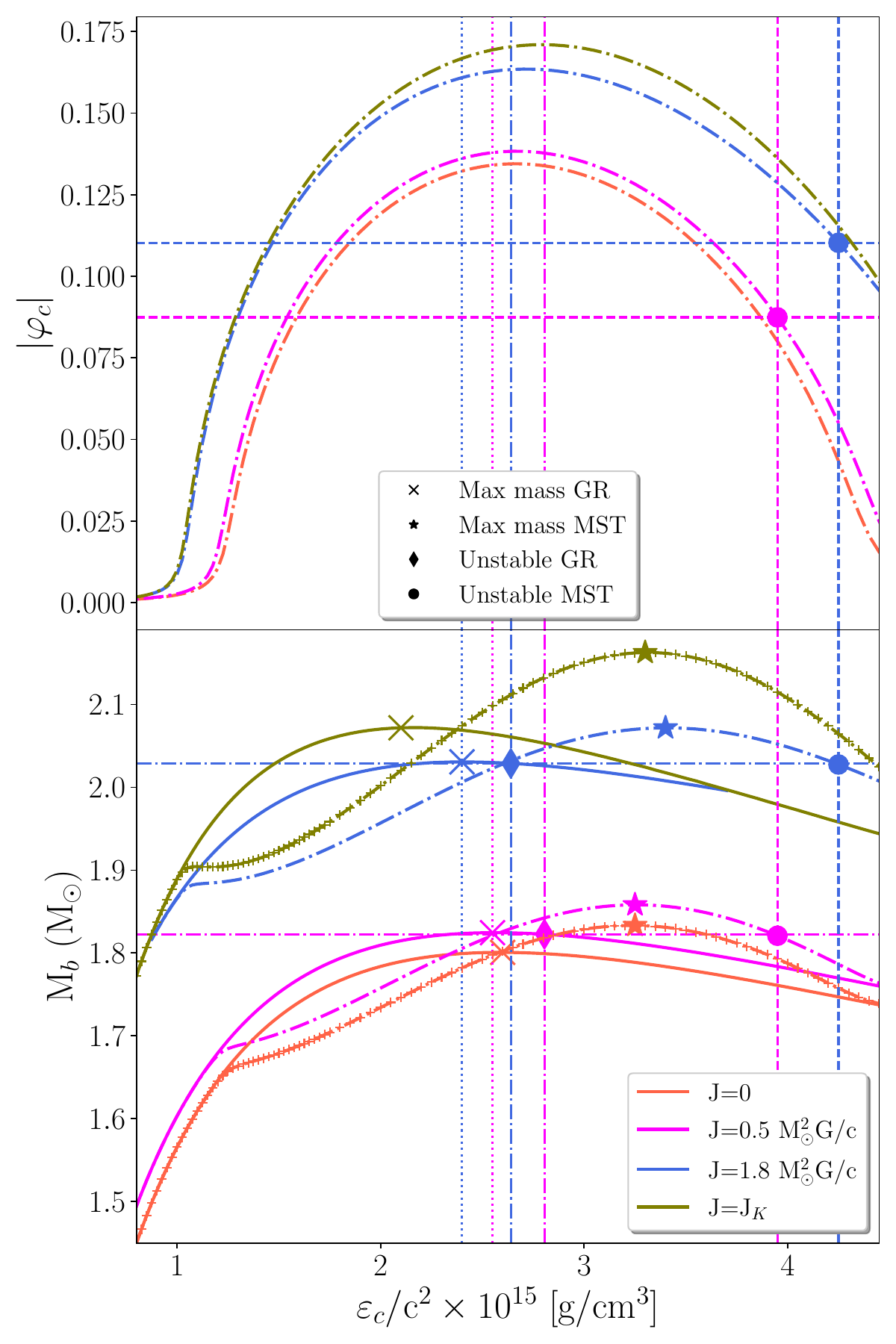}   
\caption{ The figure shows the central scalar field (top panel) and baryonic mass of the stars (bottom panel) both in GR and MSTT for different sequences of constant angular momentum. The diamonds and dots represent the GR and MSTT solutions we study on this work, whose specific details are reported in Table \ref{tab:ID_Models}. These solutions are on the same constant mass line (horizontal lines) for each constant angular momentum sequence ($J=0.5M_\odot^2,1.8 M_\odot^2$), resulting in models that have the same baryon mass and angular momentum. We also show the spherically symmetric sequence ($J=0$) and the Keplerian one ($J=J_k$) for comparison. Stars and crosses indicate the maximum mass point for each branch. }    
\label{fig:ID_full}
\end{figure}

The stellar properties for different families of constant angular momentum solutions can be seen in Fig.~\ref{fig:Mass-radius} and Fig.~\ref{fig:ID_full}. These models were computed using $\alpha_0=10^{-3}$, $\beta_0=-7.5$ and a scalar field mass of $m_\Phi=1.33\times10^{-11}$ eV/c$^2$. The figures show sequences of increasing central energy density for constant values of the angular momentum, namely $J=0.5 M_\odot^{2}$ and $1.8 M_\odot^{2}$, along with the sequence for the Keplerian limit $J_k$ and the spherically symmetric one, with $J=0$. Fig.~\ref{fig:Mass-radius} shows the magnitude of the central scalar field, the circumferential equatorial radius, the rotational frequency of the star and the spin parameter $J/M^2$. Correspondingly, Fig.~\ref{fig:ID_full} displays the gravitational mass $M$ and the central scalar field.

The first thing to notice is that the scalar field increases along with the angular momentum. That is, more rapidly rotating stars can sustain a stronger scalar field. The maximum value of the scalar field shows an increase of $\sim18\%$ between $J=0.5 M_\odot^{2}$ and $J=1.8 M_\odot^{2}$. Since rapid rotation modifies the scalar field profile, the subsequent dynamical evolution of the star will also be affected, as the degree of scalarization will also be higher than for the slowly rotating configuration. Secondly, MSTT allows for higher masses. In the case presented here, the maximum mass allowed in MSTT is $4\%$ higher than the GR case for the same angular momentum. The maximum mass star in MSTT also has a $8\%$ smaller radius than the maximum mass star in GR resulting in a more compact object. However, these values are highly dependent on the values chosen for $\alpha_0,\beta_0$ and $m_\Phi$ as well as for the EoS. Different parameter choices might result in either significant or negligible differences with respect to GR. 

For a fixed value of central energy density, a scalarized star can attain a larger angular momentum compared to GR without requiring a higher rotational frequency, reflecting how the scalar effects might also influence the properties of the spacetime and the distribution of angular momentum within the star. Since MSTT allows for higher angular momentum values than the GR configurations, this property might allow for constraining the MSTT parameters through systematic studies of collapse scenarios where the Kerr bound ($J/M^{2} > 1$) could be violated.

\vspace{.1cm}

\begin{table*}[t]
    \centering
    \caption{\textbf{Properties of the initial stellar models employed in this work.} 
    The columns report, from left to right, the name of the model, the scalar couplings $\alpha_0$ and $\beta_0$, the mass of the scalar field $m_\Phi$, the star's central rest-mass density $\rho_c$, the ratio between the polar and equatorial axis $r_p/r_e$,  the gravitational and baryonic masses $M$ and $M_b$, the equatorial circumferential radius of the star $R_e$, its rotational frequency $f$, and the ratio between the angular momentum and mass $J/M^2$.  }
    \begin{tabular}{|c|c|c|c|c|c|c|c|c|c|c|}
    \hline
      Model & $\alpha_0$ & $\beta_0$ & $m_\Phi$ [$M_\odot^{-1}$] & $\rho_c$ [$ 10^{15}$g/cm$^3$] & $r_p/r_e$ & $M \, [M_\odot]$ & $M_b [M_\odot]$ & $R_e$ [km] & f [Hz] & $J/M^2$ \\
    \hline
        GR\_S & 0 & 0 & 0 & 2.15 & 0.961 & 1.658 & 1.821 & 11.17 & 500.15 & 0.18 \\
    \hline
        MST\_S & $10^{-3}$ &  -7.5  & 0.1 & 2.73 & 0.962 & 1.658 & 1.821 & 10.53 & 556.43 & 0.18 \\ 
    \hline
        GR\_F & 0 &  0  & 0 & 2.03 & 0.673  & 1.846 & 2.027  & 13.70 & 1278.31 & 0.526 \\ 
    \hline
        MST\_F & $10^{-3}$ &  -7.5  & 0.1 & 2.89 & 0.678 & 1.845 & 2.027 & 12.50 & 1484.59 & 0.526\\
    \hline 
   
    \end{tabular}
    \label{tab:ID_Models}
\end{table*}

In this work, we focus on studying the dynamical evolution of unstable neutron stars that undergo collapse into a black hole.  For this reason, we construct constant angular momentum sequences, where the threshold of instability to collapse can be defined in terms of a critical energy density $\varepsilon_c^{\rm crit}$, as the point where $\frac{\partial M(\varepsilon_c^{\rm crit})}{\partial\varepsilon_c}|_J=0$. Stellar models with central energies higher than $\varepsilon_c^{\rm crit}$ will be unstable to axisymmetric perturbations and collapse to a black hole or migrate to a solution on the stable branch \cite{Friedman1988}. For uniformly rotating neutron stars, this turning point criterion is a sufficient, but not a necessary condition for instability, since there are models in the "stable" branch that are unstable against quasi-radial perturbations, as shown in \cite{Takami2011,Szewczyk2025}.  In the context of MSTT, the turning point criterion has been confirmed as a sufficient condition for the axisymmetric instability threshold through numerical simulations \cite{MST_rotator,Mendes2016}. For this reason, we restrict our analysis to solutions with central energy densities $\varepsilon_c>\varepsilon_c^{\rm crit}$.

Our main goal is to study the role of rotation in collapsing scalarized stars and how this process differs from GR. Thus, we consider two pairs of models. Each pair consists of a GR neutron star and a scalarized one, having the same baryonic mass and angular momentum. One of the pairs represents less massive and slowly rotating stars similar to the $D1$ model in \cite{3dcollapse1}, while the second one consists of more massive, rapidly rotating models resembling the $D4$ model in \cite{3dcollapse1}. The specifics of each model are described in Table \ref{tab:ID_Models}. 

The four models we study are also indicated in Fig.~\ref{fig:ID_full}, where we show the constant angular momentum sequences for the rapidly and slowly rotating solutions along with sequences for the non-rotating and the mass-shedding limit cases. The cross and star symbols in this figure show the maximum-mass solution for GR and MSTT, respectively. The diamond and circular markers represent the unstable solutions in GR and MSTT that we will evolve for each sequence of constant angular momentum. As can be seen from the figure, the GR and MSTT unstable solutions for each angular momentum sequence lie in the same constant mass line, resulting in configurations which have the same baryonic mass for each constant angular momentum sequence. 

\begin{figure}[h]
    \centering
\includegraphics[width=0.99\linewidth]{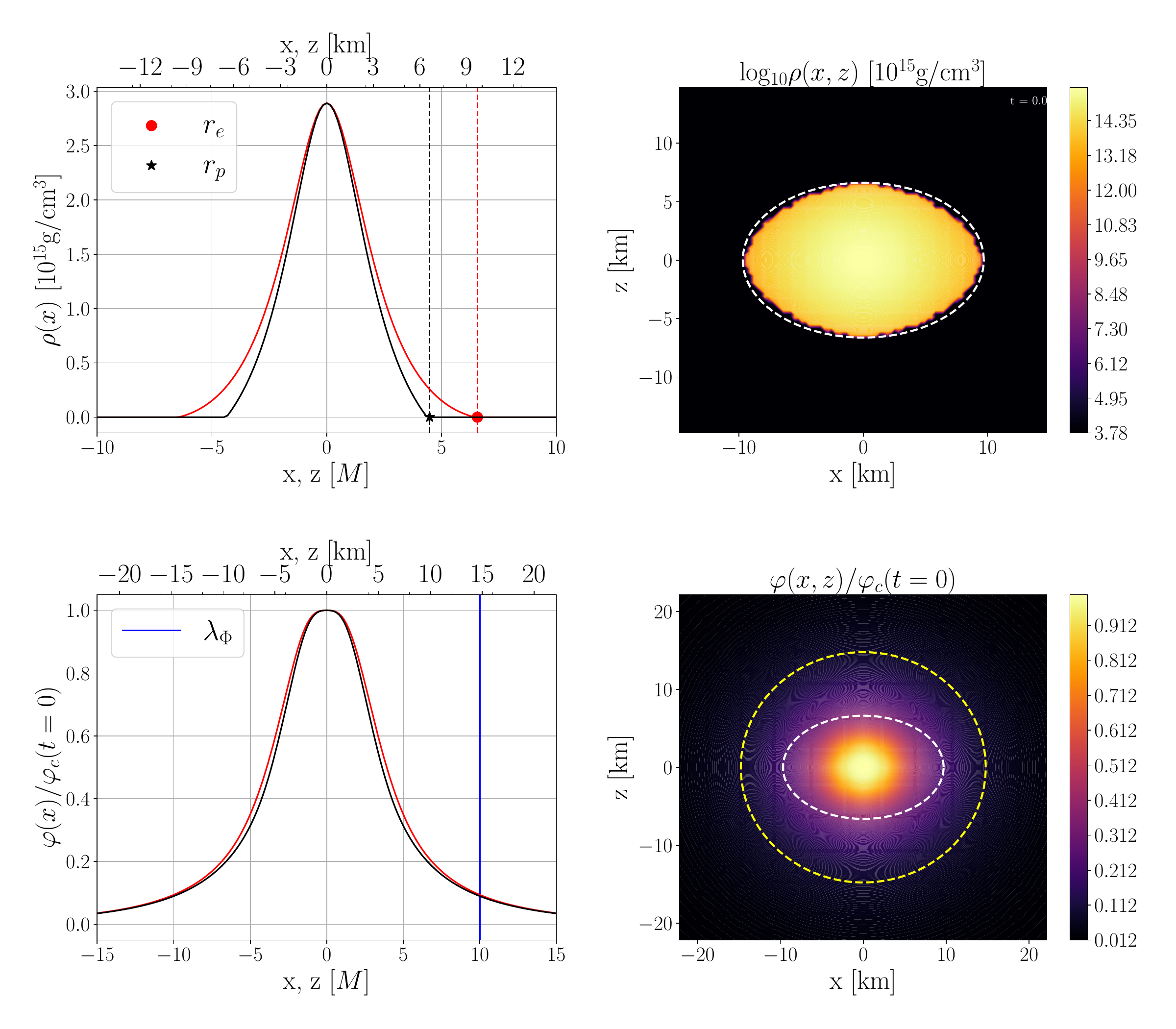}   
\caption{
The figure is organized into four panels. The top-left panel displays the initial spatial profile of the rest-mass density along the $x$- and $z$-axes (red and black lines, respectively), highlighting the polar radius (black vertical line) and equatorial radius (red vertical line). The bottom-left panel shows the corresponding profile for the scalar field, with the blue line indicating the Compton wavelength associated with the mass of the scalar field. The right panels depict the distribution of these quantities on the $x$-$z$ plane. In these views, the dashed white lines mark the stellar surface, while the dashed orange line represents the Compton wavelength. }    \label{fig:rot_ID_full}
\end{figure}

The initial configuration for a fast rotating model in MSTT can be seen in Fig.~\ref{fig:rot_ID_full}, which displays the spatial  rest-mass density distribution and the profile of the scalar field. The upper panels present the density profile of the star, where the red vertical line describes the equatorial radius $r_e$ and the black one represents the polar radius $r_p$. For this model, the ratio between the polar and equatorial radii $r_p/r_e$ is 0.678, showing the oblateness induced by rotation. The lower panels show the scalar field profile. Unlike the matter distribution, the scalar field displays less deformation along the $x$- and $z$-axes, indicating that it is less sensitive to the rotational distortion of the stellar surface. The blue vertical line marks the Compton wavelength associated with the scalar field mass. At that point, the scalar field magnitude is less than $10\%$ of its central value and starts decaying exponentially.
 
\vspace{.1cm}

%%%%%%%%%%%%%%%%%%%%%%%%%%%%%%%%%%%
\section{Numerical implementation}
%%%%%%%%%%%%%%%%%%%%%%%%%%%%%%%%%%%

\subsection{JBSSN: new thorn}

Based on the \textsc{Einstein Toolkit} infrastructure \cite{EinsteinToolkit:2025_05}, we have developed a code capable of simulating neutron stars in MSTTs. Convergence tests of the code for stable neutron star configurations are presented in Appendix B. This has been accomplished by implementing the evolution equations for the physical spacetime and the scalar field and its momentum, as shown in Eq.~\eqref{eq:SF_T}, in the BSSN formalism in the Jordan frame. 
This choice of frame allowed us to create a custom standalone spacetime evolution module (or ``thorn''), which we call \texttt{JBSSN} for Jordan frame-BSSN. Following this approach, the new thorn can be used in combination with the standard evolution thorns employed in neutron star simulations for the hydrodynamic or MHD sectors (\texttt{WhiskyTHC}, \texttt{GRHydro}, \texttt{GRHayLHD}, \texttt{IllinoisGRMHD}) without any additional modifications.  Hence, this design allows the code to be seamlessly integrated into existing workflows without disrupting the core infrastructure for matter evolution. 

In order to generate the initial data for isolated rotating stars, we employ the modified version of RNS described above, which takes advantage of the interface of the RNS version included in the \textsc{Einstein Toolkit} (thorn \texttt{HydroRNSID}) to interpolate into the three-dimensional Cartesian grid needed for the simulations. We also create an interface thorn for the scalar field, based on the \texttt{ScalarBase} thorn, part of the \texttt{Canuda} infrastructure \cite{Canuda} and expand the \texttt{ScalarInit} thorn to initialize and read the initial data for the scalar field and its momentum and interpolate the scalar field into the 3D grid. The grid settings are controlled with the \texttt{Carpet} interface \cite{Carpet2004}. 

The code employs a 4th-order finite difference scheme to calculate the spatial derivatives associated with the grid variables. The advection derivatives are calculated through a 4th-order upwind scheme with the sign of the upwind being determined by the shift $\beta$. 

For the gauge evolution equations, we employ the Bona-Masso family of slicing conditions \cite{Gauge_alpha} for the lapse, in particular the $1+$log slicing equation:
\begin{equation}
    \partial_t\alpha = \beta^i\partial_i\alpha -2\alpha K.
\end{equation}

For the shift vector, we employ the Gamma driver shift condition \cite{Gauge_beta1} through its integrated version, 
\begin{equation}
    \partial_t\beta^i= \beta^j\partial_j\beta^i+ \Gamma^i-\eta\beta^i\,,
\end{equation}
with $\eta$ acting as a damping parameter which throughout this work will take the value of $1/M$, where $M$ is the mass of the star. 

The \texttt{JBSSN} thorn is planned to be included in the \textsc{Einstein Toolkit} in the future. Through the infrastructure provided by the \textsc{Einstein Toolkit} the user can specify the  values for the coupling parameters and the mass of the scalar field, as well as the type of slicing for the lapse and the damping parameter in the shift evolution. It is also possible to select different boundary condition treatments for the scalar field depending on whether it is massive of massless (see below). This computational framework also allows to choose whether to perform the evolution using the full coupling function shown in  Eq.~(\ref{eq:Couplingfunc}), to consider the Brans-Dicke or DEF-type theories separately, or to perform evolutions in which the scalar field is fully decoupled from the spacetime and matter sectors. 
\subsubsection{Boundary conditions}

At the outer boundary of the computational grid, we assume all geometric quantities to behave as outgoing spherical waves propagating at the speed of light and we impose Sommerfeld radiation conditions like the ones discussed on \cite{boundaryAlcubierre2003} to avoid spurious reflections. Hence, each spacetime quantity at the boundary, $f(r,t)$, is expressed as,
\begin{equation}
    f(r,t) = (f_0 + u(r-vt)/r )\,,
\end{equation}
where $f_0$ is the asymptotic background value of the quantity and $u(r-v t)$ describes an outgoing wave packet traveling at velocity $v$. In this case, the velocity of the propagating wave is $v=c$. 

On the other hand, for the scalar-field-related quantities (i.e.~the scalar field and its associated momenta) we employ a Yukawa-like condition in the massive case \cite{MST_binaries_boundary}: 
\begin{equation}
    f^\varphi(r,t) = (f^\varphi_0 + u(r-v^\varphi t)/r ) e^{-mr}.
\end{equation}
Here, $v^\varphi$ represents the velocity of propagation while $f^\varphi(r,t)$ represents both the scalar field $\varphi$ and its momentum $K^\varphi$. The factor $e^{-mr}$ captures the effect of the field decaying exponentially due to the finite range interaction induced by the scalar mass, which suppresses the field amplitude beyond the Compton wavelength $\lambda_\Phi$. 

The boundary conditions are implemented in the code  terms of the differential equation, so that the right hand side is evolved. In particular, for the scalar field quantities this takes the following form, 
\begin{equation}
    \partial_t f^\varphi = -\left( v^\varphi\partial_r f^\varphi + v^\varphi(f^\varphi-f^\varphi_0)/r +v^\varphi mf^\varphi  \right).
\end{equation}
Scalar modes with different frequencies will propagate with different velocities. Along this work, as an approximation, we consider $v^\varphi\sim c$, consistent with previous studies~\cite{MST_binaries_boundary} and we have checked that this does not lead to excessive spurious incoming radiation from infinity.

\subsection{Simulation setup}

The problem of rotating neutron stars collapsing into black holes has already been extensively studied in GR, as discussed in the introduction. However, in MSTT, only spherically symmetric simulations have been performed \cite{STTcollpse}. Here, we consider the fully 3D case, following a similar approach to the methods outlined in~\cite{3dcollapse1,Collapse_Dietrich_Bernuzzi} to induce the collapse of the initial data. As stated before, we adopt the BSSN formalism for MSTT, along with the $1+$log slicing condition and the Gamma-driver condition. This choice of gauge, together with the use of adaptive mesh refinement, make it possible to maintain long-term evolutions even after the appearance of a horizon. This allows us to extract gravitational waveforms sufficiently far away from the source.

Since we are evolving the full system in the Jordan frame, the hydrodynamic equations remain the same as in GR. For this reason, we take advantage of the existing methods and implementations already available for the evolution of the fluid variables without additional modifications. The \texttt{JBSSN} thorn has been tested with hydrodynamic evolution performed using \texttt{GRHydro} \cite{GRHydro}, \texttt{IllinoisGRMHD} \cite{Illinois} and \texttt{GRHayLHD} \cite{GRHayl}, all part of the \texttt{Einstein Toolkit} infrastructure. We only present in this work the simulations performed using \texttt{GRHayLHD}. \texttt{GRHayLHD} uses the conservative formulation described in~\cite{Hydro_Ill_formulation}, along with the approximate HLLE solver \cite{Book_Riemann_SOLVER}  and the piecewise parabolic method (PPM) for reconstructing the primitive variables at the cell interfaces~\cite{PPM_method,Recovery_methods}.  For the evolutions, we employ a hybrid EoS composed of a cold polytropic part $P_{\rm cold}=K_{\rm cold}\rho^{\Gamma_{\rm cold}}$ and a thermal component $P_{\rm th}=(\Gamma_{\rm th}-1)\rho\epsilon_{\rm th}$ to model shock heating. Here, $\epsilon_{\rm th}$ denotes the thermal contribution to the specific internal energy. This hybrid formulation follows the approach by \cite{Janka1993}. We perform simulations of all the models described in Table~\ref{tab:ID_Models} using both the purely cold case and the hybrid case with $\Gamma_{th}=1.75$ (following \cite{disk_collapse}). This allows us to study whether the behavior of the additional degree of freedom (scalar field) changes when thermal support is available. In total, we perform eight simulations using the four initial data models described in Table~\ref{tab:ID_Models}. Models suffixed with \texttt{\_cold} correspond to the evolutions performed only with the cold polytrope, while \texttt{\_th} models use $\Gamma_\text{th}=1.75$.

Since we employ unstable models in hydrostatic equilibrium as initial data, the stars will eventually collapse. 
In principle, small oscillations around the equilibrium, produced by round-off truncation errors, might trigger the collapse. However, this can lead to long collapse times that depend on the grid resolution. Instead, collapse can be triggered by introducing an artificial perturbation that resembles the physical process responsible for collapse, as done e.g.~in \cite{3dcollapse1,Collapse_Dietrich_Bernuzzi}.
Physically, the mechanism driving gravitational collapse breaks the balance between the pressure gradient and gravitational binding so that gravity can take over. There are several artificial methods that can be used to break this balance. Here we consider an effective reduction of the pressure support in the core of the star. This has been previously done by decreasing the polytropic constant by a factor of $\sim0.1-2$$\%$ \cite{3dcollapse1,Collapse_Dietrich_Bernuzzi}. As mentioned in~\cite{3dcollapse2GW}, these perturbations must be small enough to avoid introducing spurious oscillations in the waveforms. Hence, in this work we keep the pressure reduction to $\Delta p/p = 0.1$. 

Technically, the Hamiltonian and momentum constraints should be solved again immediately after the introduction of this artificial pressure change. However, as discussed in~\cite{3dcollapse1}, the main dynamical properties of the star do not change by more than $1\%$, and this constraint violation (along with the violation introduced by the interpolation of a 2D solution on the 3D Cartesian grid) dissipates after a dynamical timescale of $\sim 0.2$ ms.  For all the simulations presented in this paper, the collapse begins at least after three dynamical timescales ($\sim 0.6$ ms), which allows the system to radiate away the violations in the constraints before undertaking the dynamical process. We use the \texttt{AHFinderDirect} thorn to identify and track the apparent horizon that forms during collapse. 

The simulations consist of five initial refinement levels, with the resolution at the coarsest level being $\Delta x=3.8$ km. As the star eventually collapses and a black hole horizon forms, we include three additional refinement levels, reaching a resolution on the finest grid of 0.03 km.  Each additional level is included once the lapse function crosses a threshold as it goes to zero during  black hole formation. 

Grid-based simulations of neutron stars require an artificial atmosphere in the vacuum surrounding the stars to ensure numerical stability and to prevent the simulation code from crashing.
We use a low-density atmosphere of $6\times10^{4}$ g/cm$^3$, which is around 10 orders of magnitude smaller than the initial central density for our models. Tests with different atmosphere values revealed no significant differences in the results. We also employ a 5th-order Kreiss-Oliger dissipation \cite{KO_dissipation}  on all the spacetime variables as well as in the scalar field. High values of dissipation are needed when evolving stable configurations to suppress spurious high-frequency noise which can lead to instabilities. The dissipation coefficient $\epsilon^\text{KO}$ is tuned empirically to be just large enough to suppress instabilities without affecting the star's physical properties or GW emission. We employ $\epsilon^\text{KO}\sim0.1$, which allows for long-term evolutions ($>25$ ms). We performed tests monitoring the effect induced by using significantly higher ($\epsilon^\text{KO}\sim0.6$) and lower ($\epsilon^\text{KO}\sim0.025$) dissipation coefficients. We found that for the high dissipation value, the differences with our fiducial value are negligible. In contrast, the lower dissipation case fails to suppress the high-frequency noise, triggering a numerical instability.

\subsection{GW and scalar field extraction}

To extract the emitted GWs, we compute the Newman-Penrose scalar $\Psi_4$ following the procedure outlined in~\cite{Lean} with the \texttt{NPScalars} thorn, part of the \texttt{Canuda} library \cite{Canuda}.  $\Psi_4$ is calculated by contracting the
Weyl tensor with a defined null-tetrad system. Once $\Psi_4$ has been computed over the grid, we extract its value on a set of spherical shells, where each shell is constructed at a different radius. This allows us to perform a  multipolar decomposition to obtain the individual mode contributions, which is done with the \texttt{Multipole} thorn from the \texttt{Einstein Toolkit}. This thorn interpolates the $\Psi_4$ calculated in the Cartesian grid onto spheres and then decomposes it using spin-weighted spherical harmonics. Each shell contains 120 points in the $\theta$ direction and 240 in the $\phi$ direction. 

In order to account for the energy emitted, it is also helpful to extract the scalar field which offers an additional channel of energy radiation. To extract the scalar field we also use the \texttt{Multipole} thorn. To obtain meaningful data we require observation points well outside the near-zone, far away from the source. We perform the extractions at $r_\text{ex}=(7.5\lambda_\Phi,15\lambda_\Phi,22.5\lambda_\Phi,30\lambda_\Phi)$ where the Compton wavelength is $\sim$ 15 km for the models studied in this work.
The energy associated with the scalar field can be calculated as \cite{STTcollpse}, 
\begin{equation}
    E_\Phi = \frac{c^3}{G}\int dt \left(\frac{d(r\phi)}{dt}\right)^2 \, ,
\end{equation}
where $r$ represents the extraction radius and $\phi$ is the scalar field in the Einstein frame. The squared factor represents the scalar luminosity associated with the source. 

%%%%%%%%%%%%%%%%%%
\section{Results}
%%%%%%%%%%%%%%%%%%

\subsection{Dynamical properties of the collapse}

\begin{figure}[t]
\includegraphics[width=0.9\linewidth]{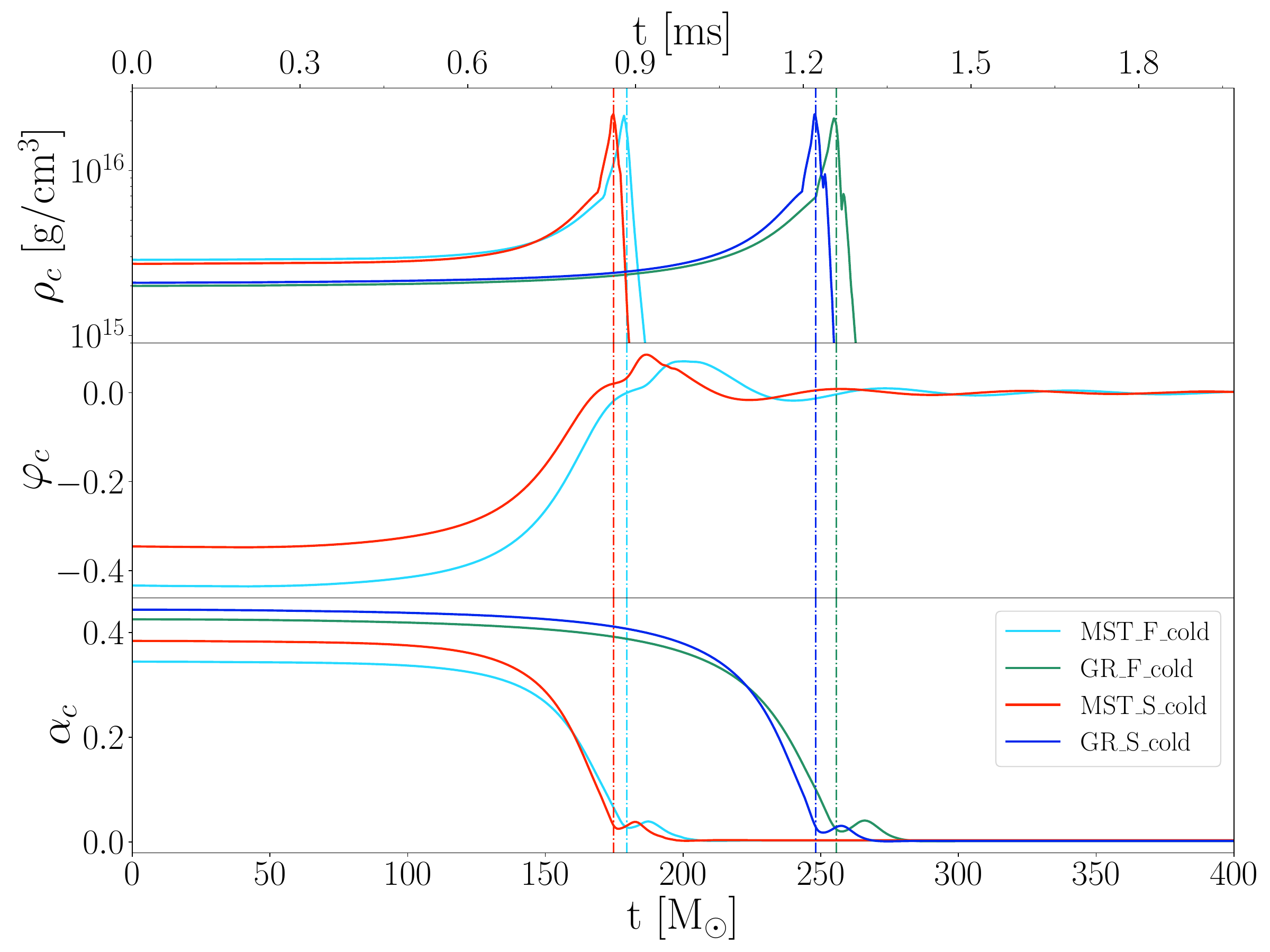}   
\caption{ Time evolution of the central rest-mass density (top panel), central scalar field (middle panel), and  of the central lapse (bottom panel). Each line is representative of the models shown in Table~\ref{tab:ID_Models} for the cold EoS evolution. Vertical lines mark the time of apparent horizon formation. }
\label{fig:timeseries_all}
\end{figure}

To discuss the dynamical simulations it is useful to first notice
an important difference between the initial stellar configurations reported in Table~\ref{tab:ID_Models}.  Even though the models have the same value of baryon mass and angular momentum in GR and MSTT for each rotation rate, the scalarized stars attain a higher central density than the ones in GR. For the rapidly rotating case, the MSTT star is almost $30\%$ denser than the GR one, while for the case of slow rotation, the central density increases approximately by $23\%$. Due to this fact, the threshold of apparent horizon formation is reached faster in MSTT than in GR, as shown by the vertical lines in Fig.~\ref{fig:timeseries_all}. Hence, an important difference in the dynamical process is the time it takes for the collapsing star to reach the maximum density.  

Fig.~\ref{fig:timeseries_all} displays the time evolution of the central values of the rest-mass density (top panel), of the scalar field (middle panel) and of the lapse function (bottom panel) for the cold EoS configurations. As the star undergoes an increase in density due to gravity progressively overcoming pressure support, an apparent horizon appears, signaling black hole formation. The horizon forms at a similar density threshold for all simulations, largely independent of the rotation rate and the values of the theory coupling parameters.  

During the collapse, the star is compressed in such a way that kinetic energy is transformed into thermal energy. The thermal adiabatic index $\Gamma_{th}$ described above is responsible for characterizing the stiffness of the thermal pressure response.  In GR, thermal pressure provides additional support against collapse, generally delaying horizon formation. In MSTT, the thermal pressure also enters the source of the scalar field equation through the trace of the stress-energy tensor $T$, coupling the thermal evolution to the dynamical descalarization of the star, with the potential to either delay or accelerate collapse. As a result, the time of horizon formation changes  when considering EoS with thermal support.

 While Fig.~\ref{fig:timeseries_all} shows only the cold case, we report the numerical values for the thermal configurations here. If the evolution is purely polytropic and cold, the horizon forms after 0.88 ms for the \texttt{MST\_F\_cold} model and 1.26 ms for its  \texttt{GR\_F\_cold} counterpart. For the slowly rotating configurations, \texttt{MST\_S\_cold} and \texttt{GR\_S\_cold}, the corresponding horizon formation times are 0.86 ms in MSTT and 1.22 ms in GR.  When considering thermal support the horizon first appears at 0.82 ms in the \texttt{MST\_F\_th} case and at 1.35 ms in the \texttt{GR\_F\_th} model. For the slowly rotating configurations, the horizon is first detected at 0.79 ms for the scalarized star \texttt{MST\_S\_th} and at 1.18 for the one in GR, \texttt{GR\_S\_th}.  

For cold EoS evolution, the collapse is on average $\sim43\%$ times faster for MSTT than in GR, with less than $2\%$ difference between the rapidly and slowly rotation configurations. This can be seen in Fig.~\ref{fig:timeseries_all}. When considering thermal effects, this difference grows to $64\%$ for the rapidly rotating case and $49\%$ for the slower rotation rate. Hence, the inclusion of thermal effects makes the difference in collapse times between GR and MSTT more pronounced, while also highlighting the dependence on the rotation rate. However, these differences on collapse times are sensitive to both the treatment of thermal pressure (different values of $\Gamma_{th}$ will change the timescale) and the initial perturbation.

\begin{figure}[t]
\includegraphics[width=0.99\linewidth]{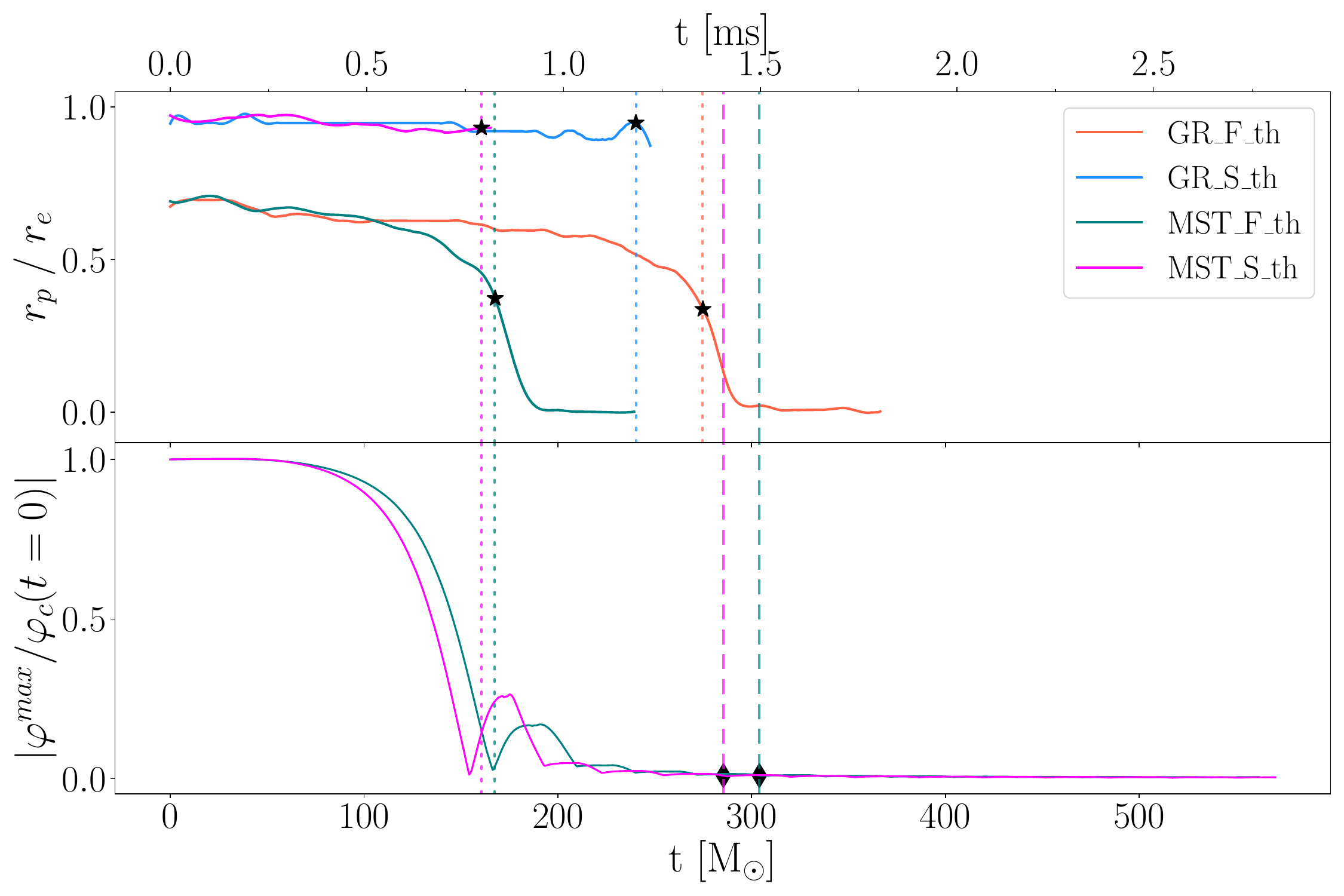}   
\caption{ (Top panel) The time evolution of the ratio between the polar and equatorial radius for all the models. The solid lines stop after all the material in the polar axis has been accreted onto the black hole. (Bottom panel) The time evolution of the maximum absolute value of the scalar field for each model. The dotted lines represent the moment the apparent horizon is first found and the dashed lines show the instant when the maximum value of the scalar field has been reduced to $1\%$ of its initial value. Note that at the time of horizon formation, the field changes sign, as can be seen from Fig.~\ref{fig:MST_fast_timeseries}.  }
\label{fig:rpre_timeseries_all}
\end{figure}

In the case of the models with slower rotation (\texttt{GR\_S\_cold, MST\_S\_cold, GR\_S\_th and MST\_S\_th}), the horizon forms earlier than for their rapidly rotating counterparts (\texttt{GR\_F\_cold, MST\_F\_cold, GR\_F\_th and MST\_F\_th}). This is expected due to the less centrifugal support against gravity in the former, an effect independent of the underlying theory of gravity. However, within a given theory, the difference in horizon formation times between slow and fast rotating models is significantly larger in MSTT than in GR.
Moreover, for the cold EoS models, the horizon appears $2.3\%$ faster for the slowly rotating model \texttt{MST\_S\_cold} than for the rapidly rotating one \texttt{MST\_F\_cold}, while in GR this difference is $3.28\%$. Correspondingly, for hybrid EoS (with thermal support), the scalarized star with slow rotation forms a horizon $3.8\%$ faster than the rapidly rotating one, while for GR configurations, this happens $14\%$ faster. One might associate this noticeable difference with the fact that MSTT stars are more compact. However, the ratio of compactness between the slow and fast rotating MSTT star is only $\sim6\%$, while it is $\sim9\%$ in the GR counterparts. Instead, the discrepancy between horizon formation times may be attributed to the stellar models having considerably higher central densities in MSTT than in GR.

\begin{figure}[h]
\includegraphics[width=0.9\linewidth]{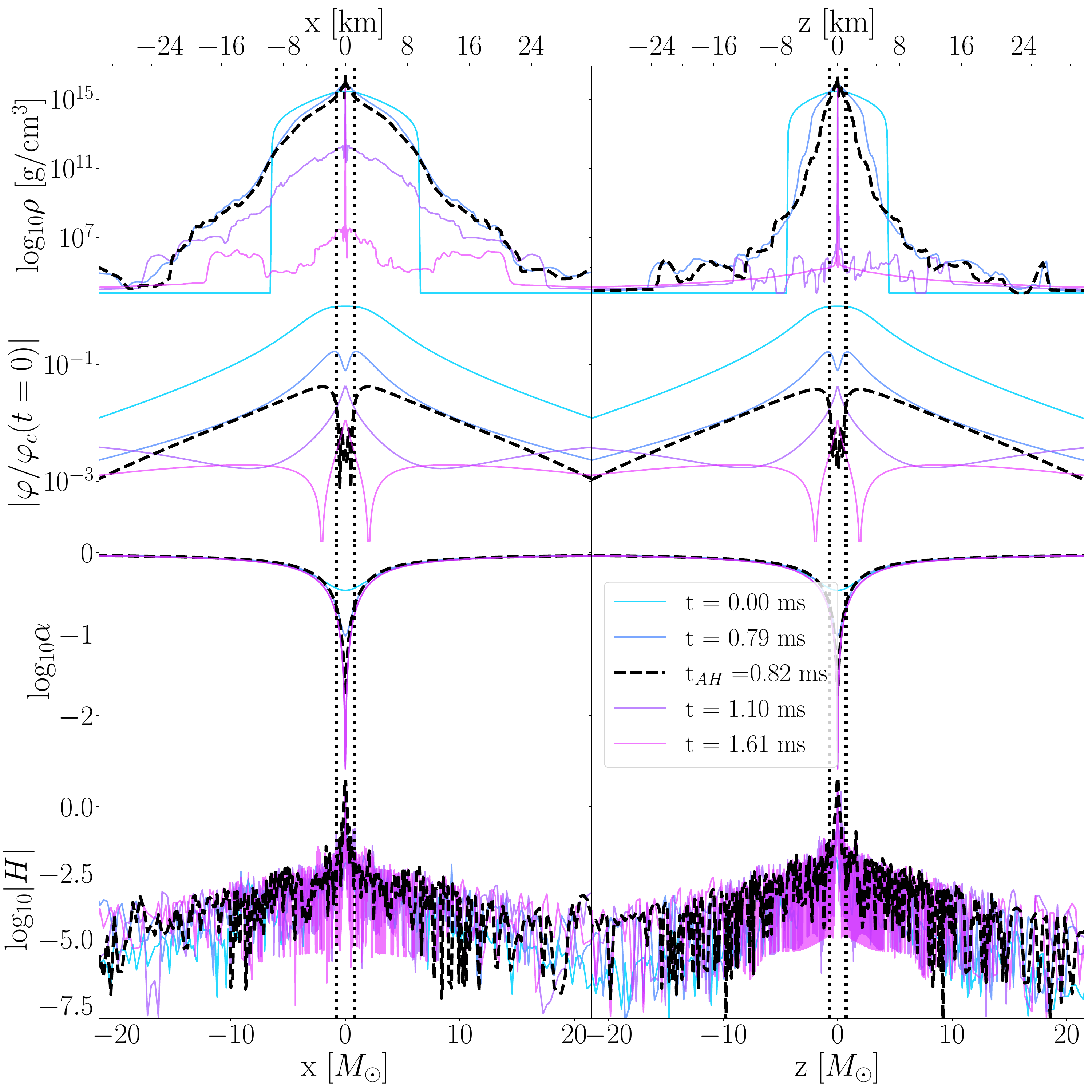}   \caption{   From top to bottom, the rest-mass density, the scalar field normalized by the initial central value, the lapse and the Hamiltonian constraint are shown.  The left panels show the $x-$profiles while the right panels indicate the $z-$profiles.  The dotted vertical lines represent the location of the horizon when it is first detected, while the dashed thick lines of the same color represent the corresponding profiles at this instant. The model shown here is \texttt{MST\_F\_th}. }
\label{fig:1dslice_fast}
\end{figure}

The top panel of Fig.~\ref{fig:rpre_timeseries_all} shows the evolution of the ratio between the polar and equatorial axes for all thermal \texttt{\_th} models (the bottom panel of this figure will be discussed later). The morphology of the star changes as the stellar surface is compressed and deformed by the inward pull of gravity. The slowly rotating models (\texttt{MST\_S\_th}, \texttt{GR\_S\_th}) collapse in an almost spherical manner. For those, the polar to equatorial deformation remains fairly constant and all the material in the polar axis is accreted by the black hole immediately after the horizon appears.  On the other hand, for the rapidly rotating solutions (\texttt{MST\_F\_th}, \texttt{GR\_F\_th}), the equatorial radius is $\sim 2.5$ times larger than the polar radius at the time of horizon formation, the star attaining a disk-like (oblate) morphology.

For those two rapidly rotating models, a transient disk forms. This is expected since the specific angular momentum of the outer edge of the star exceeds that of the innermost stable circular orbit (ISCO) of the resulting black hole, as discussed in \cite{3dcollapse1,disk_collapse}. 
This disc is short-lived as all its matter is eventually accreted by the black hole. For the cold EoS (\texttt{MST\_F\_cold}, \texttt{GR\_F\_cold}; not shown in Fig.~\ref{fig:rpre_timeseries_all}), the remaining stellar material survives for $\sim 0.21$ ms after horizon formation before being accreted by the black hole. For EoS with thermal pressure support (models \texttt{MST\_F\_th} and \texttt{GR\_F\_th}), the material is accreted more slowly, resulting in a disk that survives for around $\sim79\%$ longer than in the cold case. This is expected since the inclusion of a $\Gamma_\text{th}$ term in the hybrid EoS results in an effective stiffening of the EoS, providing additional pressure support against gravitational infall.

%--------------
Fig.~\ref{fig:1dslice_fast} displays one-dimensional spatial profiles (on the $x-$axis and $z-$axis) for the rapidly rotating model \texttt{MST\_F\_th} at different representative instants throughout the simulations. The panels show the rest-mass density (top row), the scalar field (second row), the lapse function (third row), and the Hamiltonian constraint (bottom row). The dashed lines represent the corresponding spatial profiles when the apparent horizon forms. 

The lapse at the center of the star collapses to zero as a result of black hole formation. The appearance of a disk around the black hole can also be inferred from the figure, since for the times shown after horizon formation the density in the polar axis is orders of magnitude lower than at the equatorial plane. This indicates a flattening of the stellar surface due to the rapid rotation that results in material orbiting outside of the ISCO of the resulting black hole, as mentioned above.  The Hamiltonian constraint increases sharply prior to horizon formation, particularly inside the horizon. At later times, the Hamiltonian constraint decreases for the regions outside of the horizon. This behavior is a known feature of singularity-avoiding slicing conditions, where the coordinate system becomes highly distorted near the physical singularity due to the collapse of the lapse function. Since the interior region is causally disconnected from the exterior, those large interior violations do not propagate outward and have no impact on the GW signal or on the global spacetime evolution \cite{boundaryAlcubierre2003}.  Regarding the scalar field, at the moment of apparent horizon formation the maximum value of the field has decreased to $10^{-2}$ times its initial value ($\varphi_c(t=0)$) and continues to decrease over time. During the evolution, the maximum value of the scalar field migrates from the center to the outside of the horizon, as inside the horizon all the scalar field becomes trapped. This descalarization the system undertakes will be discussed further in the following subsection. 

\begin{figure}[t]
\includegraphics[width=0.99\linewidth]{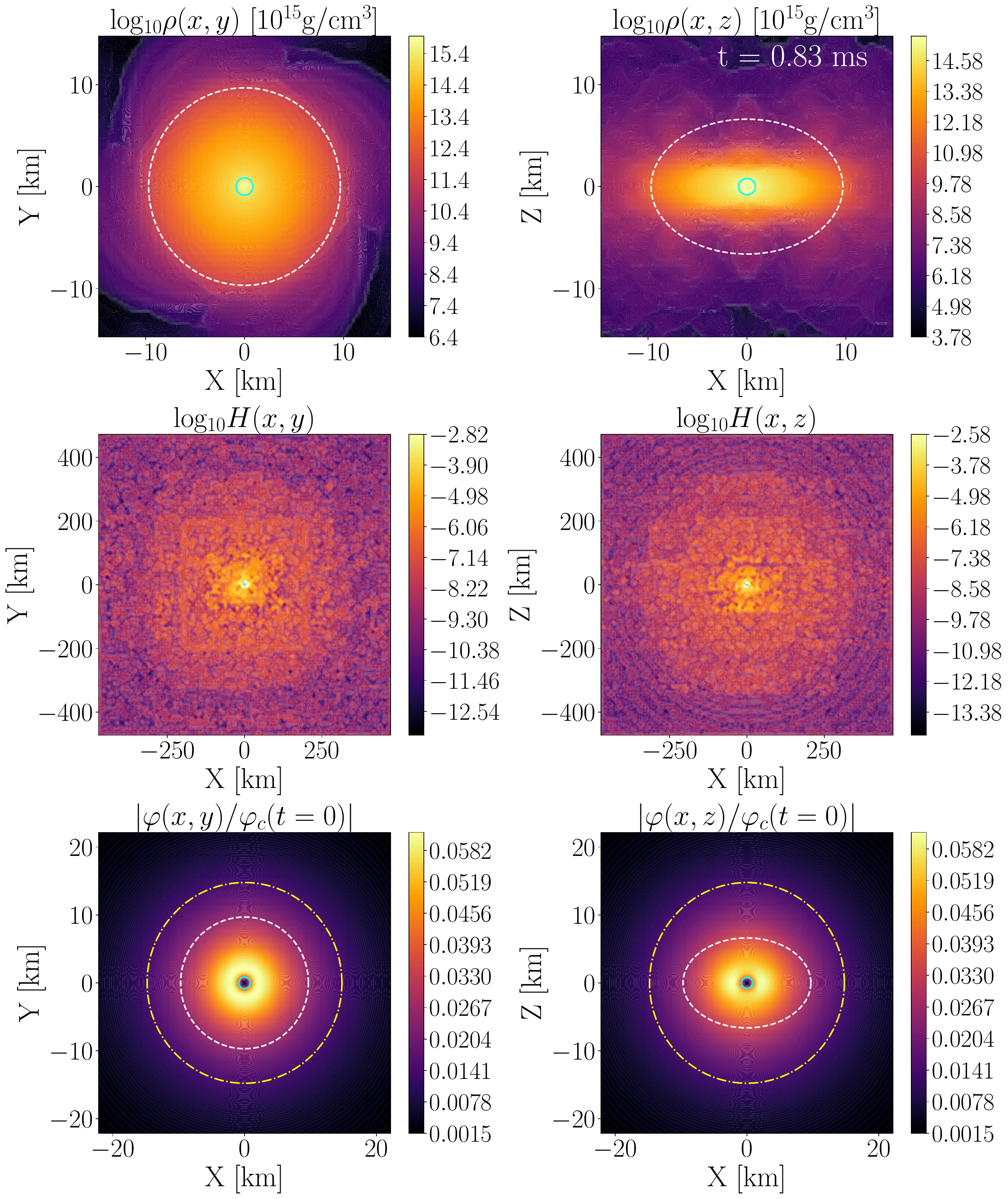}   
\caption{ Model \texttt{MST\_F\_th} immediately after horizon formation. Rest-mass density (top), Hamiltonian constraint (middle), and scalar field normalized by the central value at $t=0$ (bottom) are shown in the xy (left) and xz (right) planes. The stellar surface at $t=0$ is denoted by a white dashed line, and the apparent horizon by a solid green line. The dotted yellow line represents the Compton wavelength.   }
\label{fig:2dcollMST}
\end{figure}

\begin{figure}[t]
\includegraphics[width=0.99\linewidth]{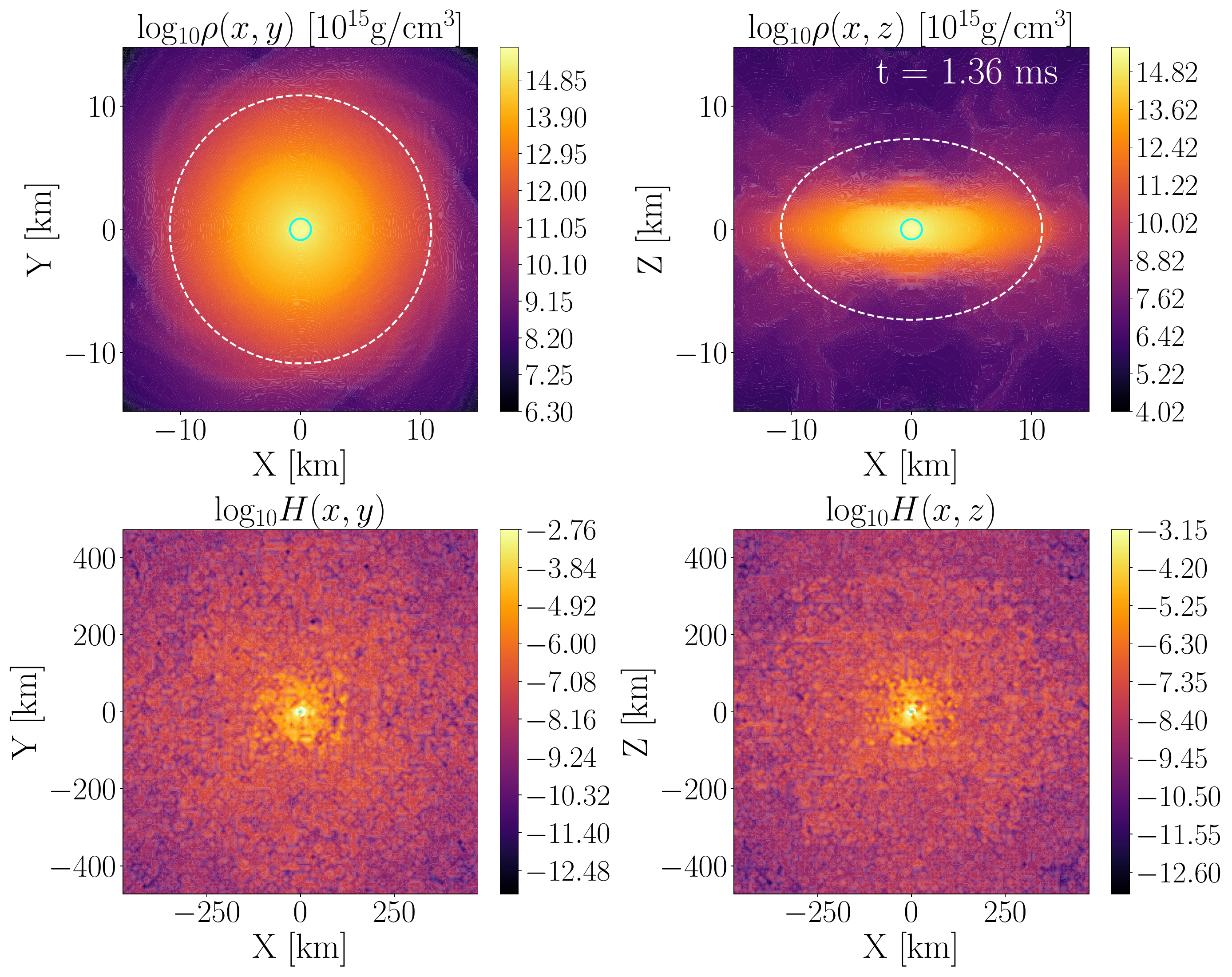}   
\caption{ Model \texttt{GR\_F\_th} immediately after horizon formation.  Rest-mass density (top) and Hamiltonian constraint (bottom) are shown in the xy (left) and xz (right) planes. The stellar surface at $t=0$ is denoted by a white dashed line, and the apparent horizon by a solid green line.} 
\label{fig:2dcollGR}
\end{figure}

Figures~\ref{fig:2dcollMST} and \ref{fig:2dcollGR} further illustrate the morphology of the collapse by depicting two-dimensional slices exactly after horizon formation for models \texttt{MST\_F\_th} and \texttt{GR\_F\_th}, respectively.
The two figures display the rest-mass density in the top panels, both in the $xy$ plane and in the $xz$ plane, with a green solid line indicating the location of the apparent horizon. The white dashed lines indicate the initial stellar surface and serve to compare the morphology of the star when the horizon forms with the initial configuration. The middle panel of  Fig.~\ref{fig:2dcollMST} shows the Hamiltonian constraint across the whole computational grid for the \texttt{MST\_F\_th} model, which can be directly compared with the GR counterpart shown in the bottom panel of Fig.~\ref{fig:2dcollGR}. 
It is worth highlighting that these panels indicate comparable magnitudes and distributions of the Hamiltonian constraint violation in both theories. This shows that there is no scalar-field-induced pathological behavior in our simulations.

In both figures (and hence for MSTT and GR) the $xy$ plane snapshots of the rest-mass density indicate that up to the instant when the apparent horizon first appears the distribution of the stellar matter remains mostly axisymmetric.  While some mass has been shed due to rapid rotation up to this point, most of the matter content remains confined within the original stellar surface. In contrast, the meridional view on the $xz$ plane in both figures 
shows that when the horizon has just formed the stars have become significantly more oblate that in the initial configurations. 
For both theories, the stellar material in the equatorial plane for rapidly rotating models remains constrained to a quasi-stable orbit. This is in contrast to the slow rotation case, where the rotational support is not enough to withstand the gravitational attraction.

Finally, the bottom panel of Fig.~\ref{fig:2dcollMST} displays the scalar field normalized by its initial central value, 
$\varphi(t)/\varphi_c(t=0)$. The yellow dashed line represents the Compton wavelength $\lambda_\Phi$ associated with the mass of the scalar field after which the field exhibits an exponential decay. By comparing with the initial configuration (see Fig.~\ref{fig:rot_ID_full}), at the time of apparent horizon formation and at $r=\lambda_\Phi$, the scalar field is already reduced to $\sim10\%$ of the central value $\varphi_c(t=0)$.     

%%%%%%%%%%%%%%%%%%%%%%%%%%%%%%%%%%%%%%%%%%
\subsection{Pre-collapse descalarization}
%%%%%%%%%%%%%%%%%%%%%%%%%%%%%%%%%%%%%%%%%%
    
\begin{figure}[t]
\includegraphics[width=0.9\linewidth]{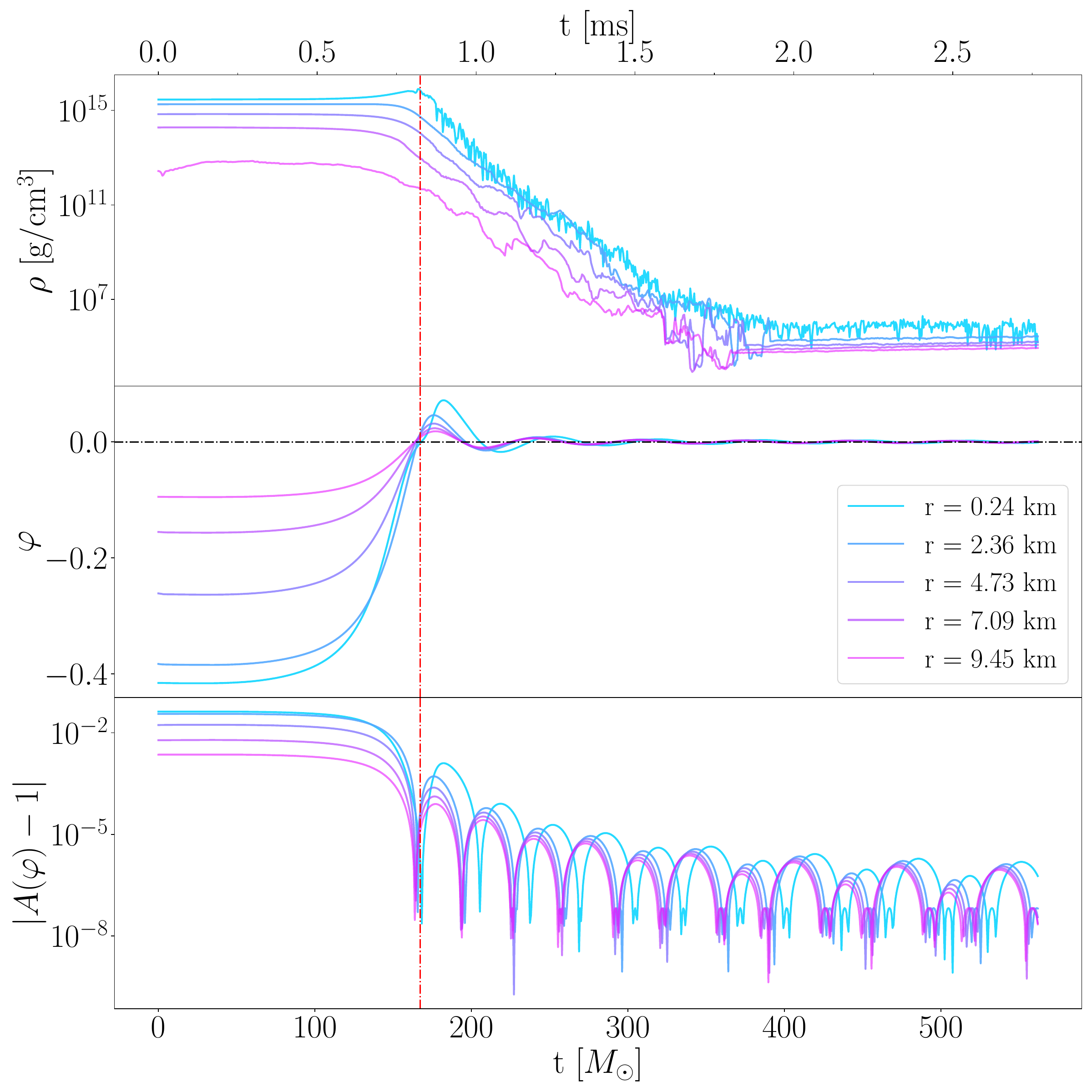}   
\caption{  Time evolution of the rest-mass density (top panel), the scalar field (middle panel) and the conformal factor (bottom panel).  The different lines represent data extracted at different radii, as shown in the legend. The red line indicates the time when the apparent horizon is formed. The model shown here is \texttt{MST\_F\_th}. }
\label{fig:MST_fast_timeseries}
\end{figure}

To understand the role the scalar field plays in the collapse it is useful to look at its time evolution as well as its spatial profiles before and after black hole formation. 

Fig.~\ref{fig:MST_fast_timeseries} displays the evolution of the rest-mass density and the scalar field for different radii for the \texttt{MST\_F\_th} model. These radii are all contained within the Compton wavelength $\lambda_\Phi$, with some of them located inside the star, whose surface is at $r_e\sim10$ km. As the central rest-mass density starts to increase, the absolute value of the scalar field in the center starts to decay. This suppression propagates quickly to the outward neighboring regions, such that by the moment of horizon formation (indicated by the vertical red line in the figure), the field is already oscillating around zero for the entire region contained within the Compton wavelength. For points outside the horizon, the scalar field is radiated away. However, because of the field being massive, a portion of the scalar field remains oscillating in the surrounding area with an amplitude of $\varphi\sim10^{-3}$. This amplitude decreases very slowly with time. This qualitative behavior is consistent across all other MSTT models we evolved and it is also consistent with the results of~\cite{MST_postmerger_remnants} for the collapse of hypermassive neutron star remnants.

Throughout the collapse, an effective potential  appears due to the strong curvature of the forming black hole. This potential barrier depends on the mass of the black hole and its spin. Moreover, the non-zero mass of the scalar field introduces a term in the Klein-Gordon equation that acts as a potential wall, creating a region where the field can be effectively trapped.
This situation resembles the long-lived scalar configurations in the form of quasi-bound states that can form dynamically around accreting black holes (see e.g.~\cite{quasi_states_scalar_in_BH}).
 
As can be seen from the spatial profiles shown in Fig.~\ref{fig:1dslice_fast}, when the apparent horizon first appears, the total magnitude of the scalar field has decreased. At that stage, the maximum of the scalar field is located outside the horizon and has decreased by more than $90\%$ with respect to its initial value $\varphi_c(t=0)$. In the subsequent evolution, most of the scalar field is radiated away, while a negligible part is trapped inside the horizon. 

This decay of the scalar field before and during horizon formation reflects a dynamical descalarization driven by the collapse itself. Our initial models sit close to the bifurcation point where the strongly scalarized branch joins the GR-like branch (see Fig.~\ref{fig:ID_full}), and along this branch the scalar amplitude decreases as the central density approaches the bifurcation. However, the dynamical evolution is not a sequence of equilibrium configurations. As the central density (scalar field) increases (decreases) rapidly due to the compression of the star, baryonic mass and angular momentum are conserved until the apparent horizon is formed. This dynamical process pushes the system into a region where no NS equilibrium with the same ($M_b$,$J$) exists and as a result, the collapse then proceeds directly into a black hole where the scalar field must decay asymptotically since the final black hole state cannot support a static scalar configuration according to the no hair theorem.

%The reason is that the central density increases while the magnitude of the scalar field decreases (along with the deformation and the change in rotation rate) drives the star away from the strongly scalarized branch into a region of central energy densities where only GR-like unstable stars exist. Thus, the scalar field is suppressed and the star migrates to its GR-like counterpart for the final stage of collapse. This pattern also appears for the slowly rotating configurations. 

%The same conclusion can be inferred directly from Fig.~\ref{fig:ID_full}. Assuming no significant angular momentum or mass loss (a reasonable assumption before the black hole is formed), the scalar field of equilibrium scalarized neutron stars decreases as the density increases. Since our choice of the scalarized solutions are already close to the bifurcation point where the strongly scalarized branch disappears and the GR-like branch regains stability, computing a stationary configuration with a central density matching that of our simulations immediately before the horizon appears yields a completely descalarized star. Naturally, such a configuration is also highly dynamically unstable because it is located beyond the maximum mass of the GR-like sequence. 

As the star descalarizes prior to the formation of the apparent horizon, the system approaches a state where $A(\varphi) \simeq 1$ and the collapse dynamics closely resembles that in GR. Therefore, since both configurations have the same baryonic mass and angular momentum, the evolution is expected to match the GR solution. However, small differences persist due to the field content trapped in the quasibound state. 
Unlike the GR case which settles rapidly to a stationary black hole, the MSTT spacetime remains dynamically active long after horizon formation as the residual field slowly radiates away.

Let us now return to discuss the bottom panel of Fig.~\ref{fig:rpre_timeseries_all} which shows in detail the maximum value of the scalar field and how it decreases over time. The dotted lines represent the moment when the horizon appears and the dashed lines show the instant when $99\%$ of the scalar field is emitted. Shortly after the horizon is formed, the maximum value of the scalar field goes to zero and peaks again as the field changes sign, as can also be noted from Fig.~\ref{fig:MST_fast_timeseries}. This is the first oscillation marking the transition between the rapid descalarization of the collapsing star and the formation of the oscillatory quasi-bound state described above.  At the moment the horizon is first detected, the maximum value of the scalar field is less than $7\%$ of the initial value and it continues to decrease as the simulation progresses. 

To contextualize the timescale it takes to radiate the scalar field, all the stellar material is accreted completely in a very short timescale of $\sim 0.1-0.4$ ms in the case of slow rotation and $\sim 0.21-1.1$ ms in the rapidly rotating case due to the disk formation.  In contrast, the time it takes for the star to suffer an almost complete descalarization (or radiating more than $99\%$ of the scalar field) is around $\sim0.6$ ms after the horizon formation (as shown in Fig.~\ref{fig:rpre_timeseries_all}). Due to the quasibound state described above, the emission process transitions from a rapid burst of scalar radiation to a slow decay in the region surrounding the black hole. The maximum scalar field value decreases to less than $0.03\%$ of the initial one at the end of the simulation.

This behavior aligns with the no-hair theorem, which requires black holes to be free of scalar hair. However, the theorem applies only to the final equilibrium state, and the spacetime remains highly dynamical after horizon formation. In addition, scalar field is ``trapped'' in a transient quasibound state that decays slowly over time due to its mass. We expect that, over significantly longer timescales, all scalar field will eventually be radiated away, allowing the system to settle into a completely hairless black hole.

%%%%%%%%%%%%%%%%%%%%%%%%%%%%%%%%%%
\subsection{Black hole formation}
%%%%%%%%%%%%%%%%%%%%%%%%%%%%%%%%%%

As discussed before, the increase in density triggers the formation of an apparent horizon, which signals the birth of a black hole. At this moment,  all the matter and scalar field content within the horizon becomes trapped and this interior region disconnects causally from the exterior. 

\begin{figure}[t]
\includegraphics[width=0.96\linewidth]{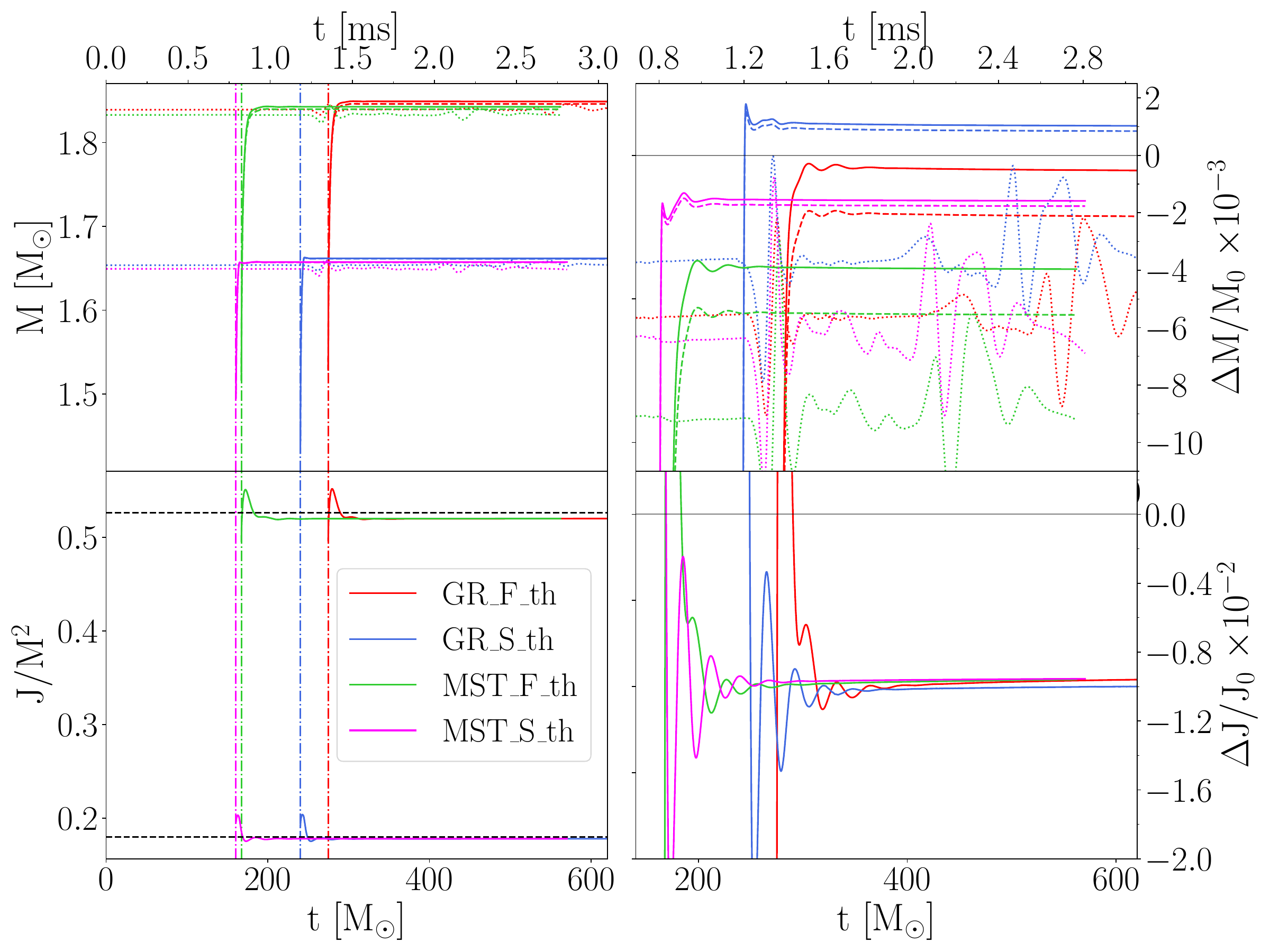}   
\caption{Top panels: evolution of the gravitational mass (left) and the relative errors with respect to the initial mass for each method (right).  Solid lines indicate the mass calculated using Christodoulou formula, dotted lines show the ADM mass computed as a surface integral at infinity, and dashed lines indicate the mass calculated through Eq.~(\ref{eq:MassCeq}). Bottom panels: evolution of the angular momentum (left) and corresponding relative error (right). Vertical lines indicate the time when the apparent horizon is first detected.}  
\label{fig:ADMmass}
\end{figure}

After the star descalarizes, the collapse in MSTTs  proceeds as it would in GR. The newly formed apparent horizon  will eventually settle to a stationary solution in which the global spacetime will resemble that of a Kerr black hole.  This can be seen in Fig.~\ref{fig:ADMmass} which shows that the total mass and angular momentum of the system remain conserved throughout the simulation, even during the descalarization phase. This implies that the scalar radiation is small in comparison with the total energy budget of the system. 

To calculate the mass of the final object displayed in Fig.~\ref{fig:ADMmass} we employ three different methods, which agree with each other up to an error of $\approx 1\%$. Those are described in Appendix D. The first method performs a standard surface integral that defines the ADM mass over the whole computational domain, as seen in Eq.~\eqref{eq:ADMmass}. This approach is agnostic to the nature of the final object and does not assume that it is a black hole. The other methods do assume that the end state is a Kerr black hole. In the second approach the total mass  is calculated by considering the geometrical properties of the horizon, whose proper equatorial circumference is  proportional to the black hole mass (cf.~Eq~\eqref{eq:MassCeq}).  The last method employs the Christodoulou mass formula for a Kerr black hole (cf.~Eq.~\eqref{eq:Christodoulou}). To use the latter, however, we need to also calculate in an accurate way the angular momentum.

This is also computed by assuming that the end state is a Kerr black hole and is based on the measurement of the distortion of the horizon. Since for a Schwarzschild black hole the horizon must be spherically symmetric, by measuring the deviation of the ratio between the polar and equatorial radius it is possible to infer the value of the angular momentum. To do this, we use Eq.~\eqref{eq:J_fit} which is a fit presented in~\cite{3dcollapse1} that parametrizes the spin parameter $a/M$ in terms of the ratio mentioned above.

The methods that rely on the assumption of a Kerr spacetime are subject to an additional systematic uncertainty associated with the fact that the spacetime is not stationary during the dynamical process. Shortly after the apparent horizon forms, the system is still subject to dynamical effects, mostly due to the accretion of matter and the quasibound state. For this reason, the approximation is mostly accurate for late times when the system has already settled to a Kerr black hole end state.   

\subsection{Gravitational wave emission}

\begin{figure}[t]
\includegraphics[width=0.9\linewidth]{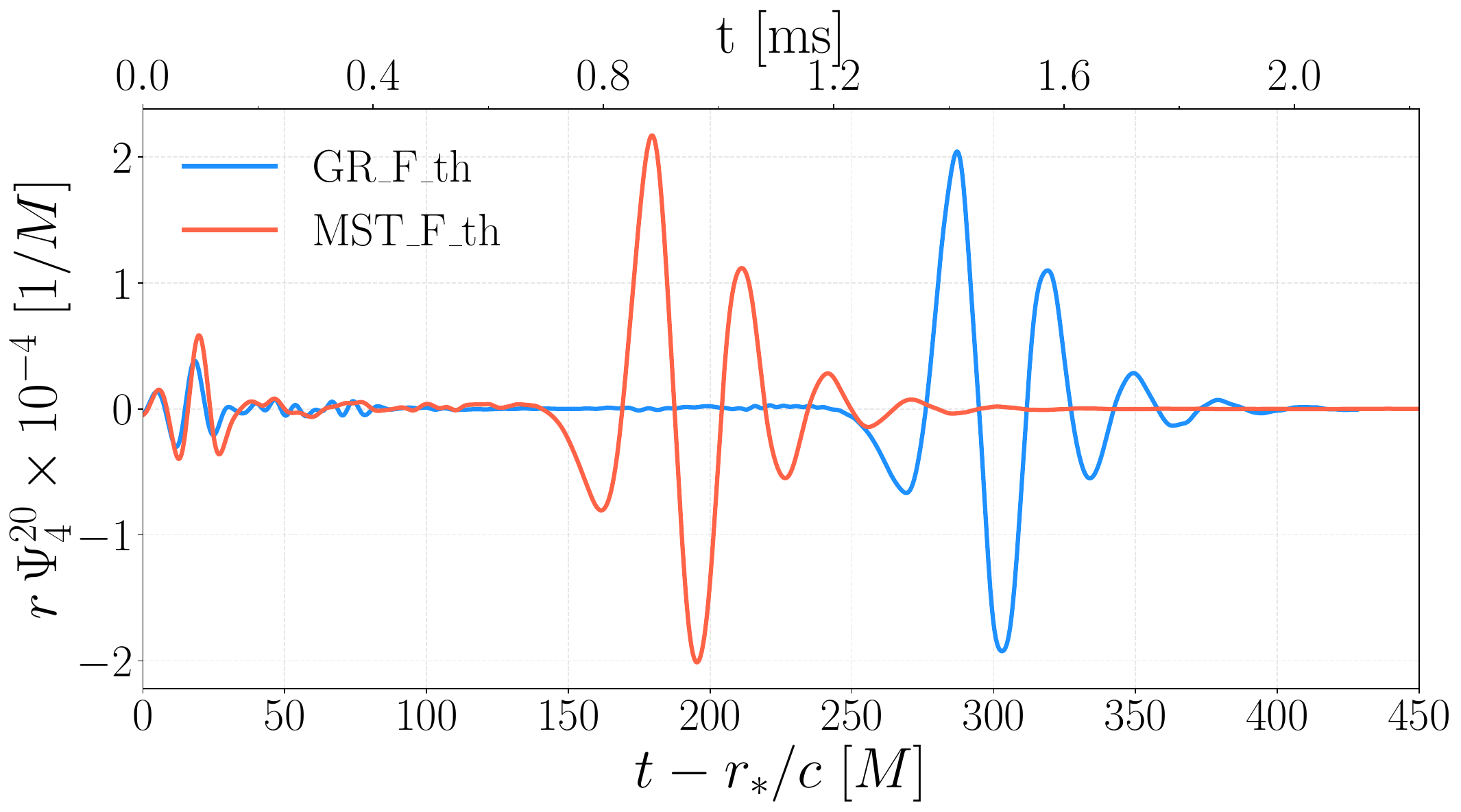}   
\caption{ $\Psi_4$ waveforms from a rotating neutron star collapsing to a black hole.  The $(l,m)=(2,0)$ mode extraction for the models \texttt{MST\_F\_th} and \texttt{GR\_F\_th}. The extraction is performed at a radius of $r_{\rm ext}=443$ km. }
\label{fig:psi4}
\end{figure}

A more complete picture of the dynamical process we study is offered by the analysis of the emitted GWs. We extract the waveforms  through the Newman-Penrose scalar $\Psi_4$. Each individual multipolar component of the radiation field is extracted by projecting $\Psi_4$ over the spin-weighted spherical harmonics associated with each specific ($l,m$) mode.
Since the collapse process is nearly axisymmetric the  gravitational radiation is dominated by the $(l,m)=(2,0)$ mode. Subdominant higher modes, such as the  $(l,m)=(4,0)$ mode, are also present. However, they are not correctly resolved by our simulation as their magnitude is similar to that of the initial perturbation introduced to trigger collapse. In particular, the amplitude of the $(l,m)=(4,0)$ mode is $\sim 50$ times smaller than that of the dominant $(l,m)=(2,0)$ mode. 
 
Fig.~\ref{fig:psi4} shows the time evolution of the $(l,m)=(2,0)$ component of $\Psi_4$ for models \texttt{MST\_F\_th} and \texttt{GR\_F\_th}, extracted at a finite radius. 
There is an initial burst of junk radiation related to the initial perturbation introduced to trigger the collapse, after which the actual signal follows. Regardless of the theory, the GW signal shows a distinctive burst morphology, with the peak of the emission  associated with the formation of the black hole. The main GW signal appears slightly earlier for MSTT compared to GR, a temporal difference which is consistent with the collapsing times discussed in previous sections. By aligning the two waveforms, it is possible to notice small signal differences in the pre-horizon phase and in the first peak after the horizon appears. However, the waveforms exhibit nearly identical evolution at late times. This is consistent with the star descalarizing, yielding late-time behavior identical to the GR case. 

\begin{figure}[t]
\includegraphics[width=0.9\linewidth]{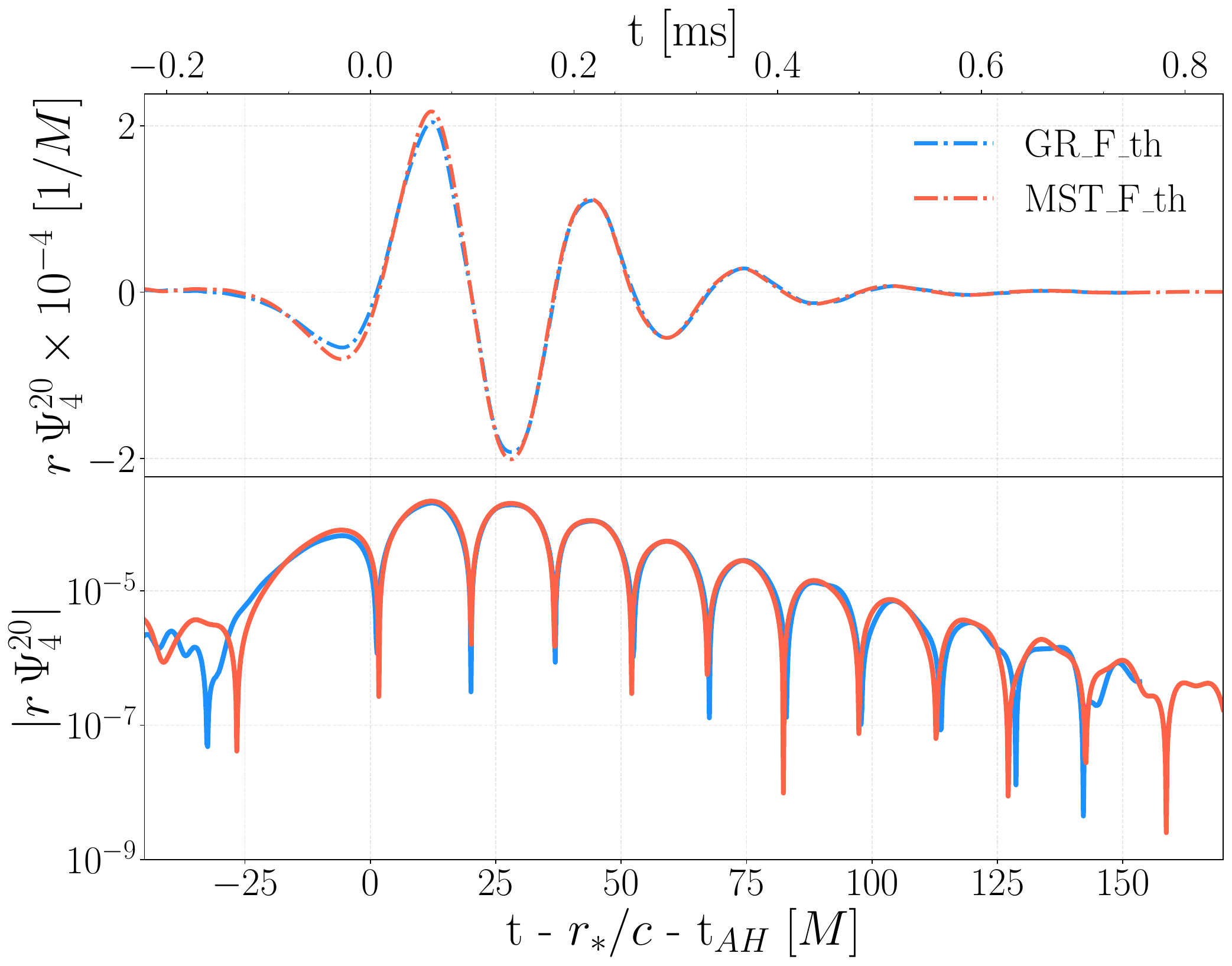}   
\caption{ $r\Psi_4$ for the dominant $(l,m)=(2,0)$ mode. The upper panel displays the waveforms aligned at the time of horizon formation. The lower panel shows the modulus in logarithmic scale to highlight the black hole ringing before settling into a Kerr black hole. Shown are the simulations for the \texttt{MST\_F\_th} and \texttt{GR\_F\_th} models.    
}
\label{fig:psi4_ring}
\end{figure}

This can be seen more clearly from Fig.~\ref{fig:psi4_ring}, which shows the waveforms associated to the $(l,m)=(2,0)$ mode of $\Psi_4$ aligned at the time of apparent horizon formation for the GR and the scalarized model.  The upper panel shows the signal aligned at the time of horizon formation, while the lower panel presents the same data in logarithmic scale to highlight the distinct quasinormal mode ringing of the final black hole. 

We also calculate the quasinormal frequencies for each case by fitting a damped sinusoid to the real and imaginary parts of $r\Psi_4$. The fitting window (same for both the GR and MSTT waveforms) spans from the moment of horizon formation to the end of the ninth oscillation cycle ($t \sim 0.7$ ms), ensuring that at least nine full cycles are captured. For the \texttt{MST\_F\_th} case, we obtain $M\omega=0.3715-0.0824i$, while the \texttt{GR\_F\_th} model yields $M\omega=0.3687-0.0798i$.  We compare these values with the theoretical quasinormal frequencies of the $(l,m)=(2,0)$ mode for a Kerr black hole with mass 1.848 $M_\odot$ and spin $a=0.526$, which we compute using the \texttt{qnm} \cite{Stein:2019mop} open-source Python package.  The GR case shows an error of $4\%$ in the real part and $7\%$ for imaginary part, while the MST case deviates $4\%$ and $5\%$ respectively. These values are consistent with those obtained in \cite{Collapse_Dietrich_Bernuzzi} for the collapse of a polytropic star in GR. However, it is well known that such errors are sensitive to the choice of the fitting window and that including higher-order overtones or modes can significantly reduce these deviations \cite{Steppohn2026}.  

\begin{figure}[t]
\includegraphics[width=0.9\linewidth]{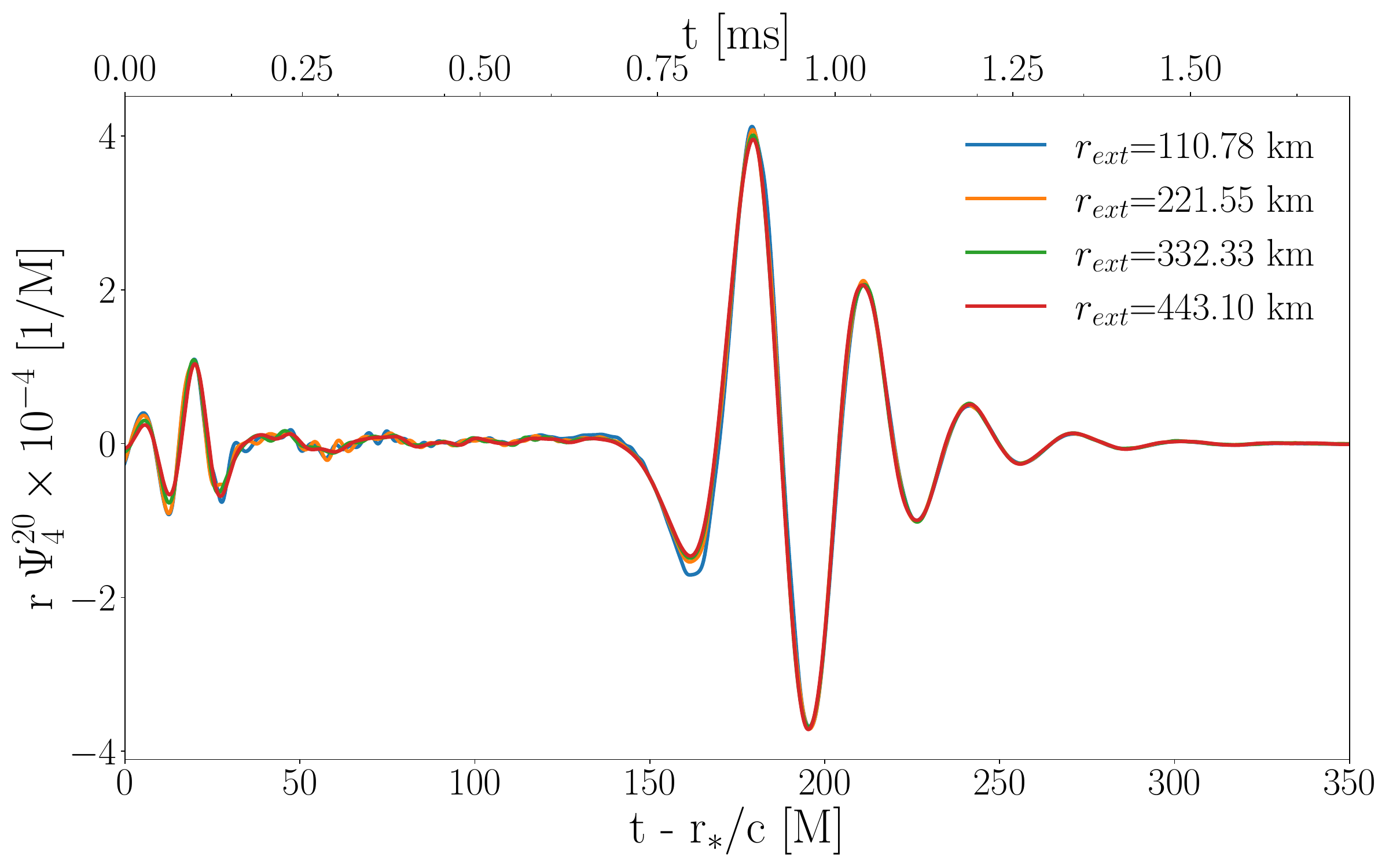}   
\caption{ The gravitational radiation from the $(l,m)=(2,0)$ mode extracted at different radii for the model MST\_F\_th.}
\label{fig:psi4_radius}
\end{figure}

As an additional diagnostic we examine the dependence of the time evolution of the emitted GWs on the extraction radius. We do this by computing the waveform at different radii and aligning the signals. We observe that a difference of 150 km between the extraction radii yields a maximum deviation in the amplitude of the relevant peaks of less than $2\%$. To avoid the errors introduced by finite-radius extraction we perform a linear extrapolation to $r\rightarrow\infty$. As can be seen from Fig.~\ref{fig:psi4_radius}, all gravitational waveforms are completely aligned when considering the retarded time for the signal, and the differences in amplitude can be seen only when the signal peaks.  

The GW strain is computed by performing a double time integration of $\Psi_4$, since $\ddot{h}^{lm}=\ddot{h}^{lm}_+ + i\ddot{h}^{lm}_\times = \Psi^{lm}_4 $.
Rather than performing the integration in the time domain, we carry out an integration over the frequency domain as described in~\cite{GW_integration}. In practice this is done through a Fourier transform of the timeseries of the multipolar components of $\Psi_4$.   

\begin{figure}[t]
\includegraphics[width=0.9\linewidth]{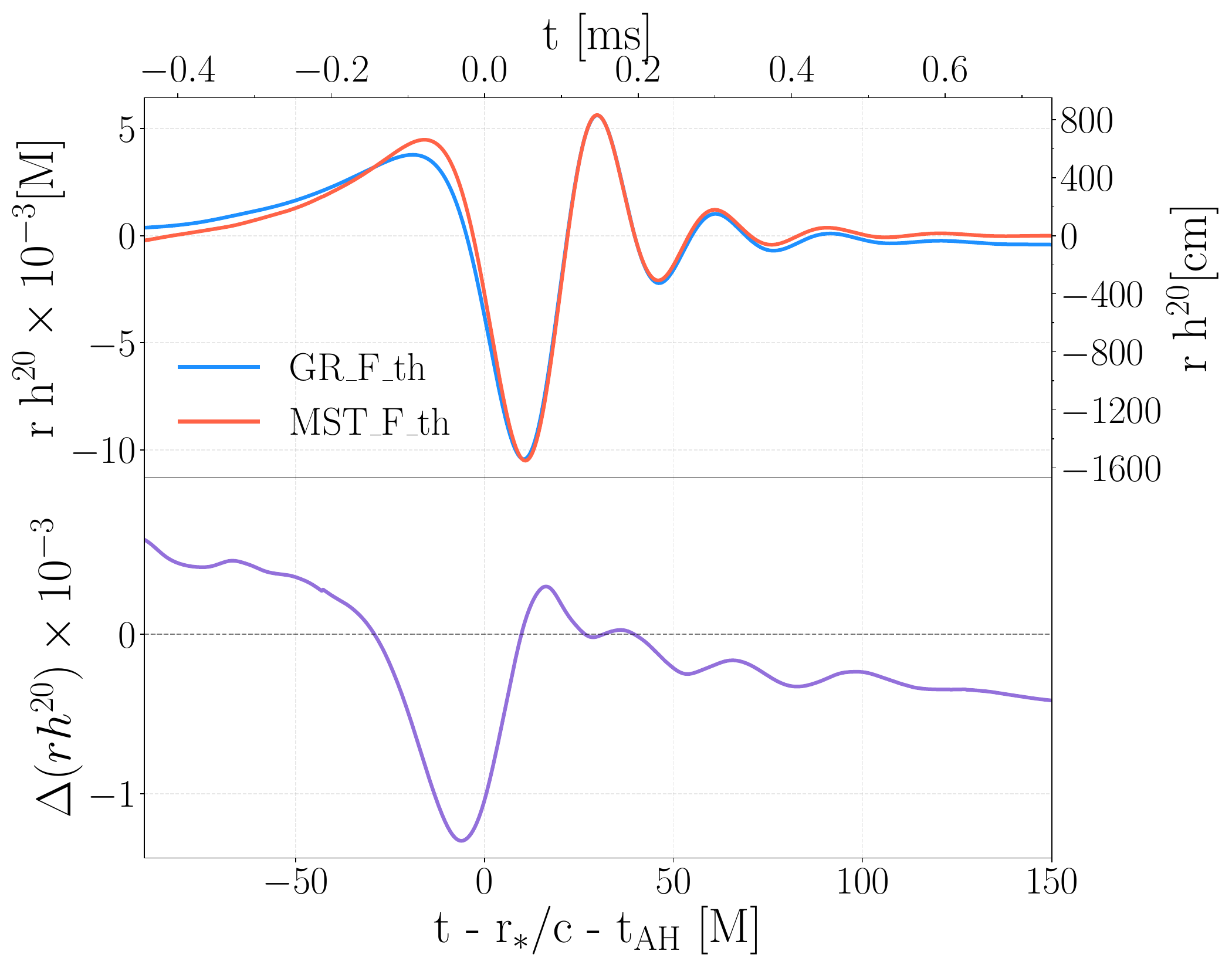}   
\caption{  Integrated waveform $rh^{lm}(t)$ for the $(l,m)=(2,0)$ mode for the \texttt{GR\_F\_th} and \texttt{MST\_F\_th} models. The top panel displays the waveforms aligned at the time of horizon formation. The lower panels shows the absolute difference between both waveforms.  }
\label{fig:h_fast}
\end{figure}

The resulting waveforms, displayed in the top panel of Fig.~\ref{fig:h_fast}, are consistent with those found in earlier works for a polytropic rotating star in GR \cite{Collapse_Dietrich_Bernuzzi,CollapseReisswig2012ThreeDimensionalGH}. Those references report directly comparable results for the well-known dynamically unstable D4 model, introduced in \cite{3dcollapse1}. Although our initial model does not correspond exactly to the D4 model, its global properties (mass and rotation rate) are fairly similar. Consequently, the waveforms computed in this paper show the same qualitative characteristics (maxima, minima, amplitude, and overall morphology) as those calculated for the D4 model in~\cite{Collapse_Dietrich_Bernuzzi}.

The bottom panel of Fig.~\ref{fig:h_fast} shows the absolute
difference between the waveform strains computed for the \texttt{GR\_F\_th} and \texttt{MST\_F\_th} models (once they have been aligned to remove artifacts induced by the initial perturbation). Although the morphology of the waveforms in both theories is similar, significant deviations are evident prior to the formation of the apparent horizon, which are a result of the descalarization dynamics as the scalar field is being emitted. These differences are significantly reduced after the horizon is formed, which suggests that the primary effect of MSTT is the energy loss associated with scalar radiation (as we discuss below), rather than the introduction of new features in the quadrupolar GW emission.

We end this section by noting that the discussed outcome is specific to black hole formation, where the no-hair theorem dictates that the scalar field is radiated away. However, if one considers the case of core collapse in MSTTs resulting in a proto-neutron star, a non-trivial scalar field may persist in the remnant, causing the GW burst to carry imprints of the scalar field dynamics. This scenario was explored in spherical symmetry with simplified EoS by \cite{scalarcollapse,corecollapse_scalar,corecollapse_scalar2} and recently in axisymmetric simulations with realistic EoS and neutrino radiation by \cite{Kuroda:2023zbz}. We intend to revisit these findings in future work.

\subsection{Scalar radiation}

We turn now to discuss the effect of rotation on the outgoing scalar radiation. For the slow rotation case (\texttt{MST\_S}), the scalar field is radiated in an almost spherically symmetric manner, with the $l=0$ mode being the most dominant. In contrast, for the rapidly rotating configurations (\texttt{MST\_F}), the scalar field also has a significant $l=2$ component, about one order of magnitude larger than for the slow rotating one but still $\sim$100 times smaller than the dominant $l=0$ mode. The $l=1$ dipole modes are also heavily suppressed. Although these modes are $10^3$ times smaller than the $l=2$ modes, they also show a considerable increase when rotation is high.
The higher-order modes are not properly resolved by the simulation, as they are of a similar order to the initial perturbation.  

\begin{figure}[th]
\includegraphics[width=0.9\linewidth]{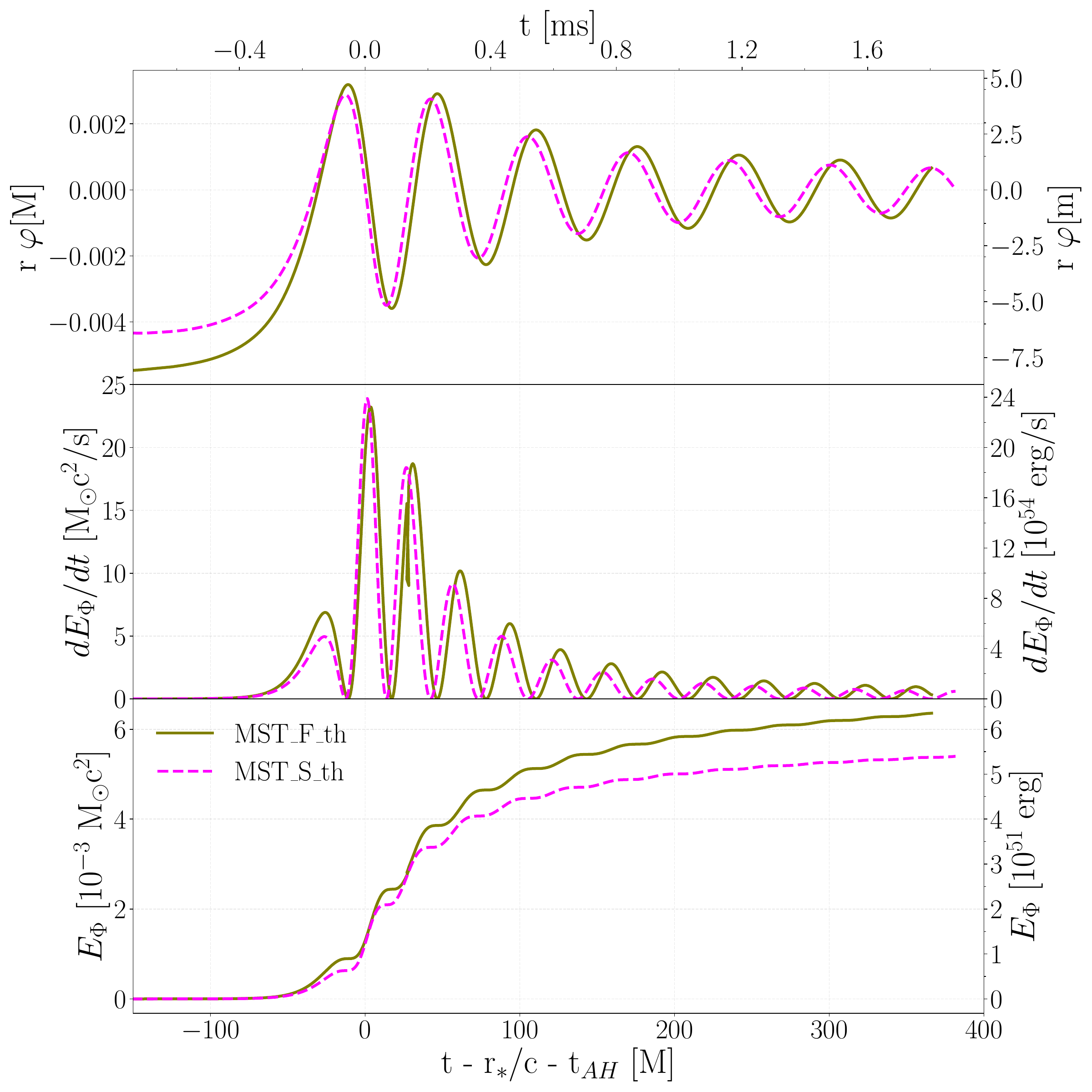}   
\caption{
Top panel: scalar field extracted at a radius of $1.5\lambda_\Phi\sim22$ km for models \texttt{MST\_F\_th} and \texttt{MST\_S\_th}. Middle panel: corresponding scalar field luminosity. Bottom panel: corresponding total energy radiated away by the scalar field.
}
\label{fig:phi_rad}
\end{figure}

Fig.~\ref{fig:phi_rad} shows that the amount of energy radiated away through the scalar field is of the same order of magnitude for both stellar models regardless of the rotation rate. This is in contrast with the quadrupolar GW emission. This radiated energy is $\sim10^{-3}M_\odot c^2$, significantly more than the energy emitted through GWs ($\sim10^{-7} M_\odot c^2$).  Even though the overall magnitude of the energy emission does not depend directly on the rotational frequency of the star, rapid rotation allows for higher scalar field amplitudes. The emission, characterized by a burst of scalar radiation, decreases faster for the slowly rotating case, with a noticeable drop in contrast to the rapid rotation case. 

This same behavior is reflected in the luminosity, which decays faster for the model with slow rotation. These differences are consistent with the underlying dynamical process, as rapid rotation provides additional support against  collapse, extending the duration of the descalarization phase. Furthermore, the rapidly rotating configuration begins with a significantly larger initial scalar field amplitude, resulting in the scalar field being radiated for an extended interval in comparison with the slowly rotating model. The rapid rotation case shows a greater energy loss, driven by this higher initial amplitude.

\begin{figure}[th]
\includegraphics[width=0.9\linewidth]{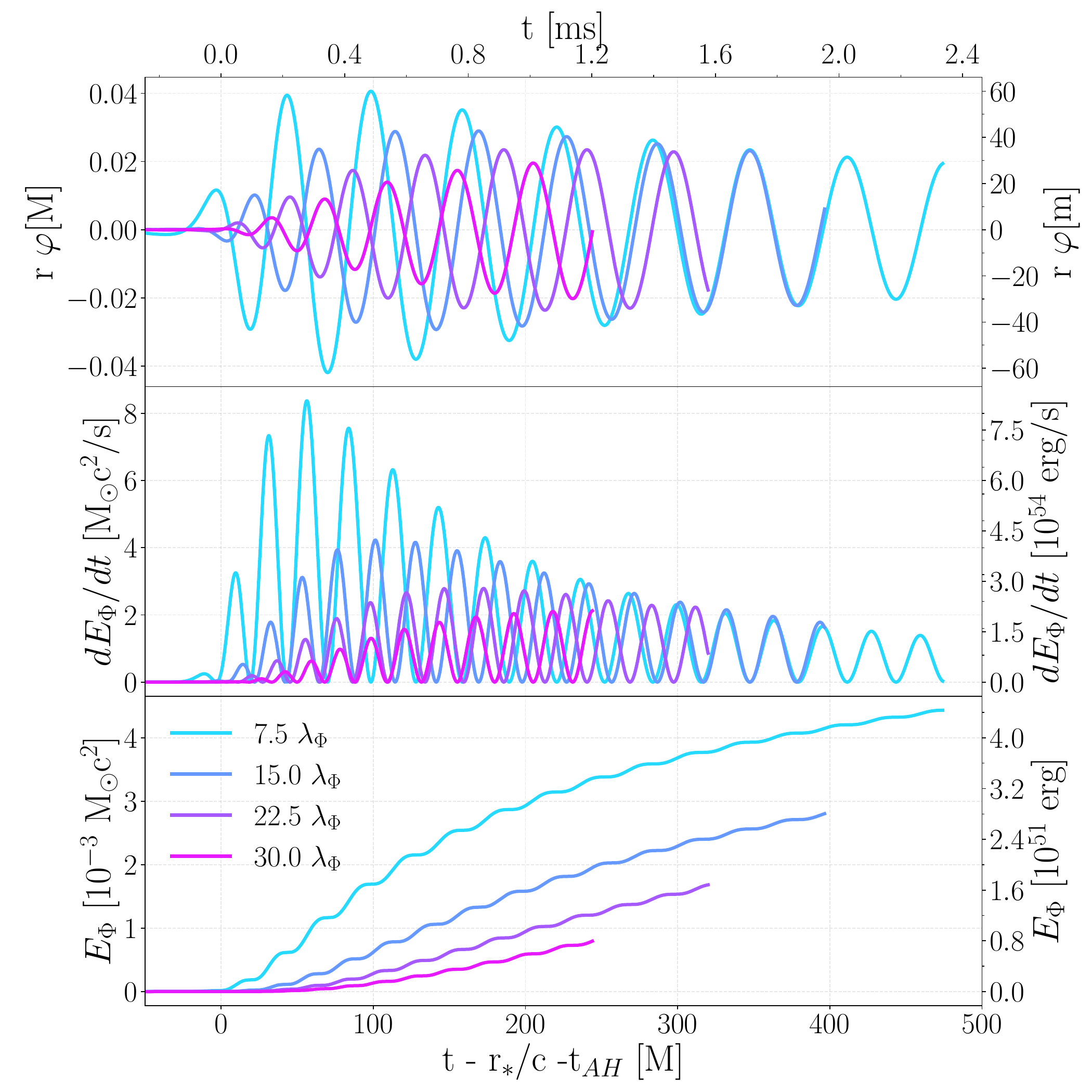}   
\caption{Top panel: scalar field extracted at different multiples of the Compton wavelength (see legend in the bottom panel) for model \texttt{MST\_F\_th}. Middle panel: corresponding scalar field luminosity. Bottom panel: corresponding total energy radiated away by the scalar field. }
\label{fig:phi_rad_fast}
\end{figure}

Fig.~\ref{fig:phi_rad_fast} shows that the energy radiated by the scalar field decreases with the extraction radius of the field. The figure displays the \texttt{MST\_F\_th} case, although the behavior is similar also for all the other scalarized models.  The different lines correspond to the signal extracted at higher multiples of the associated Compton wavelength. The curves terminate at different times because the field components arrive at the extraction zones sequentially, and the simulation ends before the full signal is captured for the farthest distances. It can also be seen that the dominant frequency in the scalar field shifts to higher values when moving further away from the star. 

%----------------

\begin{figure}[b]
\includegraphics[width=0.9\linewidth]{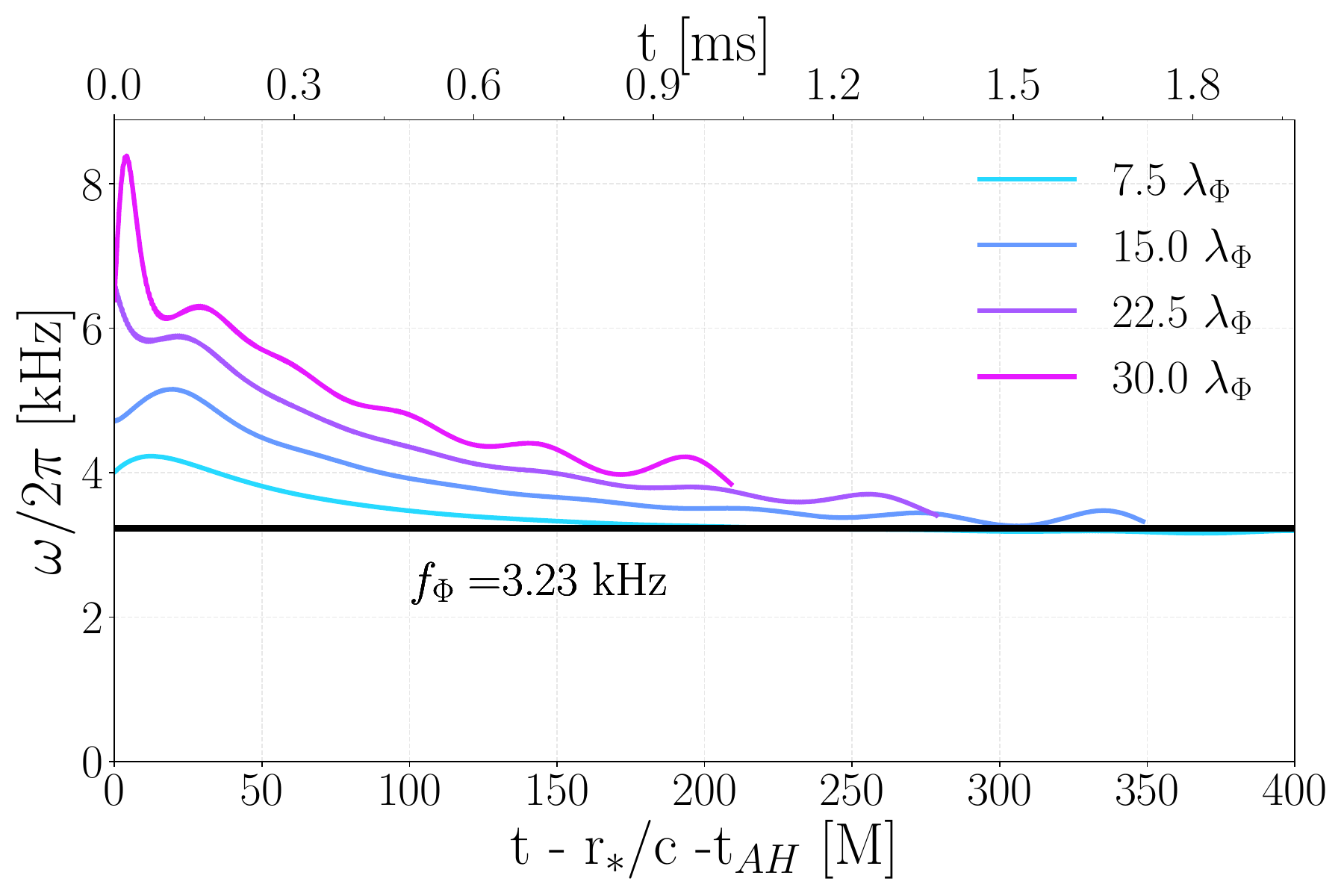}   
\caption{
Instantaneous frequency for the scalar field at the different extraction radii for the \texttt{MST\_F\_th} model. The black horizontal line shows the fundamental frequency of the scalar field $f_\Phi$. 
}
\label{fig:phi_freqs}
\end{figure}

This behavior can be seen in Fig.~\ref{fig:phi_freqs}, which displays the time evolution of the instantaneous frequency of the scalar field extractions from Fig.~\ref{fig:phi_rad_fast}. As a direct consequence of the dispersion relation followed by the scalar field, 
\begin{equation}
\omega^2=k^2+m_{\Phi}^2    
\end{equation}
each frequency mode propagates at a different group velocity and the scalar mass induces a frequency threshold such that modes with $f<f_\Phi$ are exponentially suppressed. This limit is directly derived from the mass of the scalar field $f_\Phi=m_\Phi/2\pi\sim  3.2$ kHz. As a result, the low frequency components are filtered for increasing radius, producing a spectrum that is dominated mostly by the high frequency modes and asymptotically tends to this frequency threshold.  This effect has been studied in previous works for different astrophysical processes \cite{descalarization,scalarcollapse,corecollapse_scalar}. 

This dispersive behavior is also reflected in the scalar luminosity, which decreases with distance as fewer modes reach the extraction surface, resulting in a lower radiated energy, as expected. The qualitative trends for the luminosity and the radiated energy are consistent with previous studies of accretion induced descalarization \cite{descalarization}.

Another consequence of attributing mass to the scalar field becomes apparent when comparing with previous results for the massless case, such as \cite{STTcollpse}, where simulations of spherically symmetric collapsing scalarized neutron stars were discussed. In the massless case, regardless of the coupling parameters $\alpha_0$ and $\beta_0$, the quantity $r\varphi$ shows a smooth, monotonic decay to zero during the descalarization phase, where the oscillations we see in Fig.~\ref{fig:phi_rad} are absent (see figures 6 to 9 from \cite{STTcollpse}). 

In our case, the inclusion of a massive term induces an oscillatory behavior around zero which persists both in the extracted monopolar waveform and in the long lived quasibound state discussed above. After sufficient time, the frequency of these oscillations matches the theoretical prediction coming from $m_\Phi$, as can be seen from Fig.~\ref{fig:phi_freqs}. This behavior appears even if the scalar field is extracted inside the Compton wavelength, where the notion of a scalar charge still exists. However, unlike in the massless case, the quantity $r\phi$ is no longer a constant at infinity in the massive case. Since for $r>\lambda_\Phi$, $r\phi\sim e^{-r/\lambda_\Phi}$, the notion of a conserved scalar charge is no longer well defined in the radiation zone.

The oscillatory structure described above is a direct signature of the massive Klein-Gordon dynamics, which are absent in the massless limit where the field propagates strictly at the speed of light without a frequency dependence.   In the radiation zone, the massive field obeys a dispersive relation where the group velocity $v_g$ for each frequency mode is different. As the wave packet propagates outward, lower-frequency components travel slower than higher-frequency ones, causing the signal to stretch temporally. This dispersion results in a frequency-dependent amplitude profile at asymptotic distances, a behavior consistent with the findings in \cite{corecollapse_scalar}.

%%%%%%%%%%%%%%%%%%%%%
\section{Conclusions}
%%%%%%%%%%%%%%%%%%%%%

In this paper we have investigated the role of rotation in the collapse of neutron stars to black holes within Massive Scalar-Tensor Theories (MSTTs). To this end, we have developed an extension of the \textsc{Einstein Toolkit} capable of simulating such gravitational collapse in full 3D within MSTTs\footnote{Our spacetime evolution module (or ``thorn''), dubbed \texttt{JBSSN} for Jordan frame-BSSN, will soon be made  publicly available as a set of computational modules for the \textsc{Einstein Toolkit}.}. The implementation of the evolution equations, based on a modified version of the BSSN formalism, has been done in the Jordan frame representation for a general form of scalar field coupling function including the Brans-Dicke and Darmour-Esposito-Farese models, in the case of nonzero scalar field mass, following~\cite{Shibata-2014}.   Our new code has successfully simulated the gravitational dynamics of matter and spacetime in the presence of a massive scalar field coupled to the geometric sector. We have simulated uniformly rotating, scalarized neutron stars and compared their collapse with GR counterparts, focusing on the global dynamical properties of the process, the formation and characterization of the resulting black hole, and the emission of both GWs and scalar radiation. This work represents a first step toward full 3D simulations of rotating core collapse in MSTT. In this initial study, we have adopted a simple polytropic EoS, deferring to future work the analysis of more realistic neutron star matter.

We have found that the main effect of MSTT appears on the collapse timescale, although this might be partially a consequence of two factors: the impact of the numerical perturbation on  the initial data to trigger the collapse and the intrinsically higher central densities of the scalarized solutions compared with their GR counterparts, which place those configurations closer to the collapse threshold.  {The qualitative features of the dynamical evolution remain similar in MSTTs and in GR. The most notable differences appear due to the additional channel of energy loss through scalar radiation emitted during the descalarization phase. The total amount of energy radiated is directly dependent on the initial magnitude of the scalar field, and indirectly dependent on the angular momentum of the star, since more rapidly rotating configurations support a higher value of scalar field. 
Faster collapse (signalled by the appearance of an apparent horizon) in MSTTs happens independently of the rotation rate; however, for stars with lower angular momentum, the collapse happens even faster due to insufficient centrifugal support to delay it.

The end product of the collapse is consistent with the formation of a Kerr black hole in both GR and MSTT. All simulations we have performed agree, within numerical errors, with the expected result regarding a collapsed object characterized by the mass and angular momentum of a given Kerr spacetime. The associated GW emission is characterized by a distinctive burst-like signal. 
Our results show that the main morphological features of the waveforms are fairly similar in MSST and in GR. We see a different observational signature coming from the descalarization phase on the early pre-horizon section of the waveforms. However, this effect is subdominant, since the signal peaks right after the horizon forms, where the dynamical properties of the collapsing star are not strongly influenced by the scalar field, with hydrodynamical effects being dominant. This might change for other matter models less simplistic than polytropes. In particular, \cite{Shao_scalar_core} have shown that when considering realistic EoS multiple scalarized solutions might appear for the same energy density, which could have an effect on the GW emission.  

Our results suggest that GW observations from the gravitational collapse of rotating neutron stars might not be able to distinguish between  extended MSTTs and GR by using tensor modes alone. The only channel that can potentially break the observational degeneracy between collapsing scalarized stars and their GR counterparts is the emission of scalar radiation. Most of the information associated with the scalar field is radiated away through the scalar component of the emission, a type of radiation not present in GR and that can help constrain the mass of the scalar field and the coupling parameters.

In this work we have also studied the role of rotation on such scalar radiation. For slowly rotating configurations, the scalar luminosity decays more rapidly, reflecting a faster descalarization of the star. In these cases, the scalar emission is initially stronger than in rapidly rotating stars. Hence, a comparatively larger amount of energy is released at early times of the collapse. 
As a result, the total scalar content of a slowly rotating star is depleted more quickly, resulting in a reduced luminosity at later stages.
In contrast, the rapidly rotating configurations initially support a higher value of the scalar field amplitude, and the descalarization process is slightly delayed. This spreads the emission over a longer temporal window, resulting in a more gradual decay of the luminosity profile. The detectability prospects of scalar radition from neutron star collapse will be reported elsewhere.\\

\section*{Acknowledgments}
This work is supported by the Spanish Agencia Estatal de Investigación (grant PID2024-159689NB-C21) funded by MICIU/AEI/10.13039/501100011033 and by FEDER / EU, by the Generalitat Valenciana (Prometeo Excellence Programme grant CIPROM/2022/49), and by the European Horizon Europe staff exchange programme HORIZON-MSCA2021-SE-01 (grant NewFunFiCO-101086251). JCOM and DD acknowledge financial support via an Emmy Noether Research Group funded by the German Research
Foundation (DFG) under Grant No. DO 1771/1-1. JCOM acknowledges partial support by Secretaria de Ciencias, Humanidades y Tecnologías de México (SECIHTI). DD acknowledges financial support by the Spanish Ministry of Science, Innovation, and Universities via the Ram\'on y Cajal programme (grant RYC2023-042559-I), funded by MICIU/AEI/10.13039/501100011033 and by ESF+. The partial support of KP-06-N62/6 from the Bulgarian science fund is also gratefully acknowledged. The authors acknowledge support by the High Performance and Cloud Computing Group at the Zentrum für Datenverarbeitung of the University of Tübingen, the state of Baden-Württemberg through bwHPC and the German Research Foundation (DFG) through grant no. INST 37/935-1 FUGG. 
We acknowledge Discoverer PetaSC and EuroHPC JU for awarding this project access to Discoverer supercomputer resources. 

%%%% Bibliography %%%%%
\bibliographystyle{apsrev4-2}
\bibliography{bib}

\appendix
\numberwithin{equation}{section}

\section{Evolution equations for MSTT in Jordan frame}
\subsection{The BSSN equations}
We consider a spacetime foliation consisting of a family of spacelike hypersurfaces normal to a timelike unit vector $n^\mu$. This foliation is characterized by the lapse function $\alpha$ and the shift vector $\beta^i$, where each hypersurface is described by the 3-dimensional metric $\gamma_{ab}$. 
 Considering a conformal metric $\bar{\gamma}_{ij}$ which is related to the 3-metric through a conformal factor $W$, $\bar{\gamma}_{ij}=W^2\gamma_{ij}$,  and separating the curvature tensor in terms of its trace $K$ and its traceless components $A_{ij}$ as $K_{ij}=A_{ij}+\frac{1}{3}\gamma_{ij}K$, one can obtain:

  \begin{align}
\partial_t \tilde{\gamma}_{ij} &= \beta^k \partial_k \tilde{\gamma}_{ij}-2\alpha \tilde{A}_{ij} + 2 \tilde{\gamma}_{k(j} \partial_{i)} \beta^k - \frac{2}{3} \tilde{\gamma}_{ij} \partial_k \beta^k, 
\end{align}
\begin{align}
\partial_t W &= \beta^k \partial_k W+ \frac{1}{3} W \left(\alpha K -\beta^k\partial_k\right), 
\end{align}
\begin{equation}
\begin{aligned}
\partial_t K &= \beta^k\partial_kK-D^k\partial_k\alpha +\alpha\left(\tilde{A}^{ij}\tilde{A}_{ij}+\frac{1}{3}K^2\right)\\ &+\frac{4\pi\alpha}{\Phi}(\rho+S) 
+\frac{\alpha\omega(\Phi)}{\Phi^2}\Pi_\Phi^2\\&+\frac{\alpha}{\Phi}\left[ \tilde{\gamma}^{ij}W^2D_i\partial_j\Phi-k\Pi_\Phi-\frac{3}{2}F(\Phi)  \right]\\
\end{aligned}
\end{equation}
\begin{equation}
\begin{aligned}
    \partial_t \tilde{A}_{ij} &= \beta^k\partial_k\tilde{A}_{ij} +2 \tilde{A}{k(i} \partial_{j)} \beta^k - \frac{2}{3} \tilde{A}_{ij}\partial_k\beta^k \\&+W^2 \left( -D_i\partial_j \alpha + \alpha \tilde{R}{ij} \right)^{TF} \\
&+ \alpha \left( -2\tilde{A}{ik} \tilde{A}_j^k +  K\tilde{A}_{ij} \right)  - \frac{\alpha\omega(\Phi)}{\Phi^2} W^2 (\partial_i\Phi \partial_j\Phi)^{TF}  \\
& -\frac{\alpha}{\Phi}\left( W^2 D_i(\partial_j\Phi)^{TF}-\tilde{A}_{ij}\Pi_\Phi \right)   -8\pi\frac{ \alpha}{\Phi}  W^2 S_{ij}^{TF},
\end{aligned}
\end{equation}

\begin{equation}
    \begin{aligned}
        \partial_t \tilde{\Gamma}^i &= \beta^k\partial_k\tilde{\Gamma}^i - \tilde{\Gamma}^k\partial_k\beta^i + \frac{2}{3} \tilde{\Gamma}^{i} \partial_k \beta^k \\ &- 2 \alpha \tilde{A}^{ij} \left( \frac{3}{W}\partial_jW +\partial_j\alpha \right)
+ 2 \alpha\tilde{A}^{jk} \tilde{\Gamma}_{jk}^i \\& +\tilde{\gamma}^{jk} \partial_j\partial_k \beta^i + \frac{1}{3} \tilde{\gamma}^{ik} \partial_k \partial_j \beta^j \\ &-\frac{4}{3}\alpha\tilde{\gamma}^{ij}\partial_jK -16\pi\alpha\frac{J^i}{\Phi}  -2\alpha\frac{\omega(\Phi)}{\Phi^2} \Pi_\Phi\tilde{\gamma}^{ij}\partial_j\Phi \\& - \frac{2\alpha}{\Phi}\left[\tilde{\gamma}^{ij}\left(\partial_j\Pi_\Phi-\frac{1}{3}K\partial_j\Phi   \right) -\tilde{A}^{ij}\partial_j\Phi\right],
    \end{aligned}
\end{equation}

where the function $F(\Phi)$ is defined as:  
\begin{equation}
\begin{aligned}
&F(\Phi)=\frac{1}{2\omega(\Phi)+3}\left[ 8\pi T +\Phi^3\frac{d}{d\Phi} \left(\frac{U(\Phi)}{\Phi^2} \right) \right. \\ & \left. +\frac{d\omega}{d\Phi}\left( \Pi_\Phi^2-W^2\tilde{\gamma}^{ij}\partial_i\Phi\partial_j\Phi \right)  \right]
\end{aligned}
\end{equation}

\subsection{Constraints}

The Hamiltonian and momentum constraint change with respect to GR through the source terms. The Hamiltonian constraint is given by
\begin{equation}
    \begin{aligned}
            \mathcal{H} = R + K^2 - (A_{kl}A^{kl}+\frac{1}{3}K^2) -\frac{16\pi}{\Phi}\left( \rho_M + \rho^\Phi\right)\\,
    \end{aligned}
\end{equation}

with
\begin{equation}
    \begin{aligned}
        \rho^\Phi = \frac{\omega(\Phi)}{16\pi\Phi}\left(\Pi_\Phi^2 + \partial_k \Phi \partial^k \Phi\right)\\+\frac{1} {8\pi}\left( -K\Pi_\Phi+D_k(\partial^k\Phi) + \frac{1}{2}U(\Phi)   \right).
    \end{aligned}
\end{equation}

The momentum constraint can be defined as

\begin{equation}
        \mathcal{M} = D_kA_{i}^k-\frac{2}{3}\partial_iK -\frac{8\pi}{\Phi}(J^M_i+J_i^\Phi )
\end{equation}

with 
\begin{equation}
    J_i^\Phi=\frac{\omega(\Phi)}{8\pi\Phi}\Pi_\Phi\partial_i\Phi + \frac{1}{8\pi}\left( \partial_i \Pi_\Phi -A^j_i\partial_j \Phi -\frac{1}{3}K\partial_i\Phi\right)
\end{equation}

\section{Convergence and validation of the code}

In this appendix, we present a convergence analysis of our code. In order to test the stability and accuracy of the code we perform evolutions of stationary stable neutron star solutions in MSTT. We examine two scenarios, a non-rotating configuration and a rapidly rotating equilibrium model. The solutions considered here are $20\%$ more massive than their GR counterparts for the same central rest-mass density. The grid employed in this set of simulations extends up to 128 $M_\odot$ ($\sim$ 189 km) with 4 refinement levels, the innermost located at 14.77 km. The grid spacing in the innermost refinement level is 0.22, 0.295 and 0.369 km for the high, medium and low resolution simulations, respectively.

\begin{figure}[h]
\includegraphics[width=0.975\linewidth]{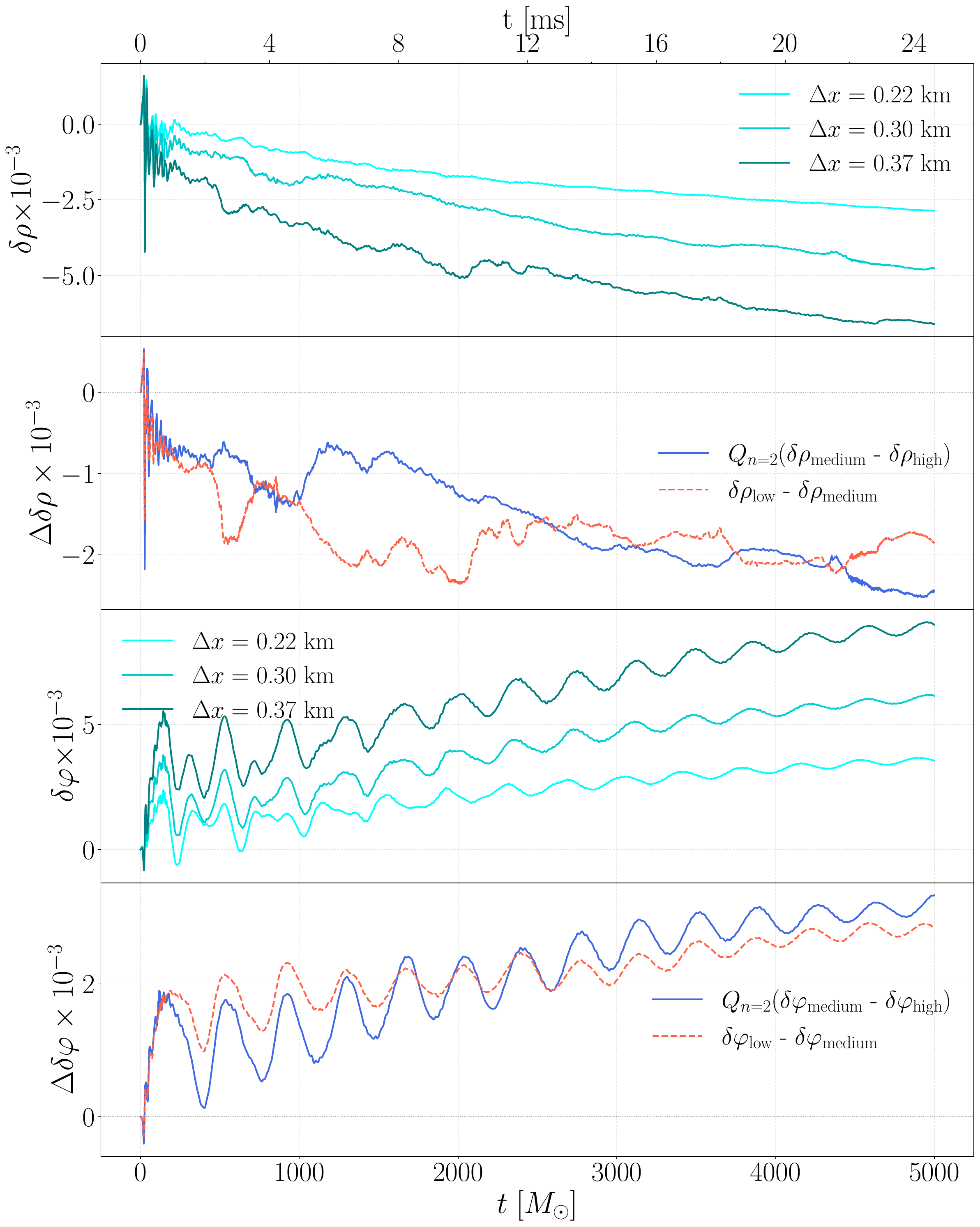}   
\caption{ Central rest mass density and scalar field evolution for a spherically symmetric star with gravitational mass of $1.7$ $M_\odot$. The scalar field is determined by a mass of $7.5\times10^{-2}\, M_\odot^{-1}$, with $\alpha_0=0.001$ and $\beta_0=-7$. The upper panel shows the evolution of the relative error with respect to the initial data for the central density, while the second row panel shows the convergence of different resolutions. The third row panel shows the evolution of the error for the central scalar field and the last panel shows the convergence of different resolutions.
The grid spacing at the finest refinement level is 0.22, 0.295 and 0.369 km for the high, medium and low resolution runs respectively. }
\label{fig:convergenceTOV}
\end{figure}

For an $n-$th order of convergence, we calculate the factor
\begin{equation}
    Q_n= \frac{\Delta x_{\mathrm{low}}^n-\Delta x_{\mathrm{medium}}^n}{\Delta x_{\mathrm{medium}}^n-\Delta x_{\mathrm{high}}^n}
\end{equation} 
for the three resolutions $\Delta x$ described above.

We define for a variable f, the relative error $\delta f= f(t)/f(t=0)-1$ and the difference between two runs using different resolutions A and B as $\Delta f_{A-B}$. For ideal $n$th order convergence, $Q_n = \Delta f_\mathrm{low-medium}/\Delta f_\mathrm{medium-high}$.  To validate our code, we monitor the relative error in the central density $\delta\rho$ and the central scalar field $\delta\varphi$ with respect to the initial data for the spherically symmetric and rapidly rotating configurations mentioned above.

\begin{figure}[h]
\includegraphics[width=0.975\linewidth]{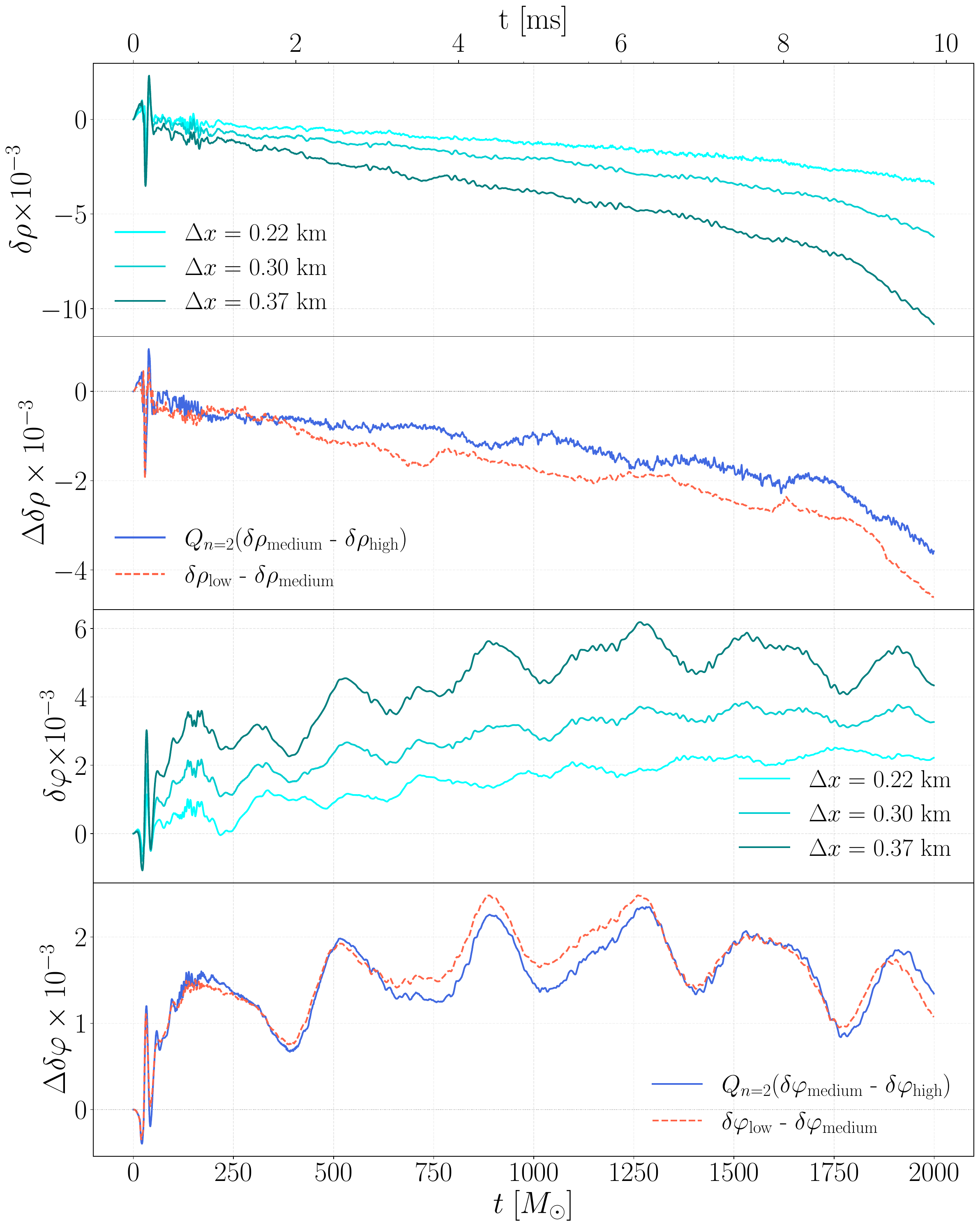}   
\caption{Central density and scalar field evolution for a rapidly rotating star with gravitational mass of $2.2725$ $M_\odot$ rotating at $\Omega=0.95\Omega_k$. The mass of the scalar field is of  $10^{-1}\,M_\odot^{-1}$, with $\alpha_0=0.001$ and $\beta_0=-8$. The same panel description and resolutions as in Fig.~\ref{fig:convergenceTOV}. }
\label{fig:convergenceRot}
\end{figure}

The equilibrium solution for the spherically symmetric case is modeled with a piecewise polytropic approximation \cite{piecewisepoly2009} of the SLy \cite{Sly4} EoS. The star has a mass of 1.8892 $M_\odot$, a radius of 11.56 km and a central rest-mass density of 1.26 $\times10^{15}$ g/cm$^3$. The mass of the scalar field is $5\times10^{-2}\, M_\odot^{-1}$, which corresponds to a Compton wavelength of $\sim$30 km, while the coupling parameters are $\beta_0=-7$ and $\alpha_0=10^{-3}$. 

The evolution of the spherically symmetric star can be seen in Fig.~\ref{fig:convergenceTOV}, where we display the time evolution for $\delta\rho$ on the first row panel and $\delta\varphi$  on the third. The second and fourth row panels show the difference for low and medium resolution along with the difference for high and medium resolutions scaled by the ratio $Q$ expected for second-order convergence. The relative errors with respect to our initial data are comparable to those presented in ref. \cite{Shibata-2014} for the massless case.

For the rapidly rotating configuration, we use a piecewise polytropic parametrization of the APR EoS \cite{APR4}. The resulting star has a mass of 2.27 $M_\odot$, a circumferential equatorial radius of 13 km, and is rotating with a spin parameter of $J/M^2=$0.56. In this case, the scalar field mass has a value of $10^{-1}\, M_\odot^{-1}$ with a Compton wavelength of 14.7 km. The coupling parameters are $\alpha_0=10^{-3}$ and $\beta_0=-8$. The results can be seen in Fig.~\ref{fig:convergenceRot}. The layout of this figure is the same as for the spherically symmetric case in Fig~\ref{fig:convergenceTOV}.

In order to validate the code in a dynamical regime, we perform simulations of a star undergoing spontaneous scalarization starting from a GR-equilibrium initial configuration. The star has a gravitational mass of 2.45$M_\odot$ and is rotating slowly at a frequency of 0.6 Hz. We assume a scalar field mass of $10^{-1}\,M_\odot^{-1}$, with $\alpha_0=0.001$ and $\beta_0=-12$. The results can be seen in Fig.~\ref{fig:convergence_scalarizarion}. The panels, arranged from top to bottom, display the evolution of the central rest-mass density, the Hamiltonian constraint, the scalar field, and the lapse function, as well as the relative error in the conserved baryonic mass. For this test, we employ the DD2 EoS \cite{DD2} in the finite temperature version from \cite{EOS_Oconnor2010}, which is publicly available at \cite{stellarcollapse}.  We employ three different resolutions, 0.24 (high), 0.295 (medium) and 0.369 (low) km. We observe second-order convergence for the scalar field, metric variables, and hydrodynamic quantities. The relative error in the baryonic mass also exhibits second-order convergence, remaining below $0.02\%$ throughout the simulation (up to t=20 ms for the lowest resolution).

\begin{figure}[hb]
\includegraphics[width=0.975\linewidth]{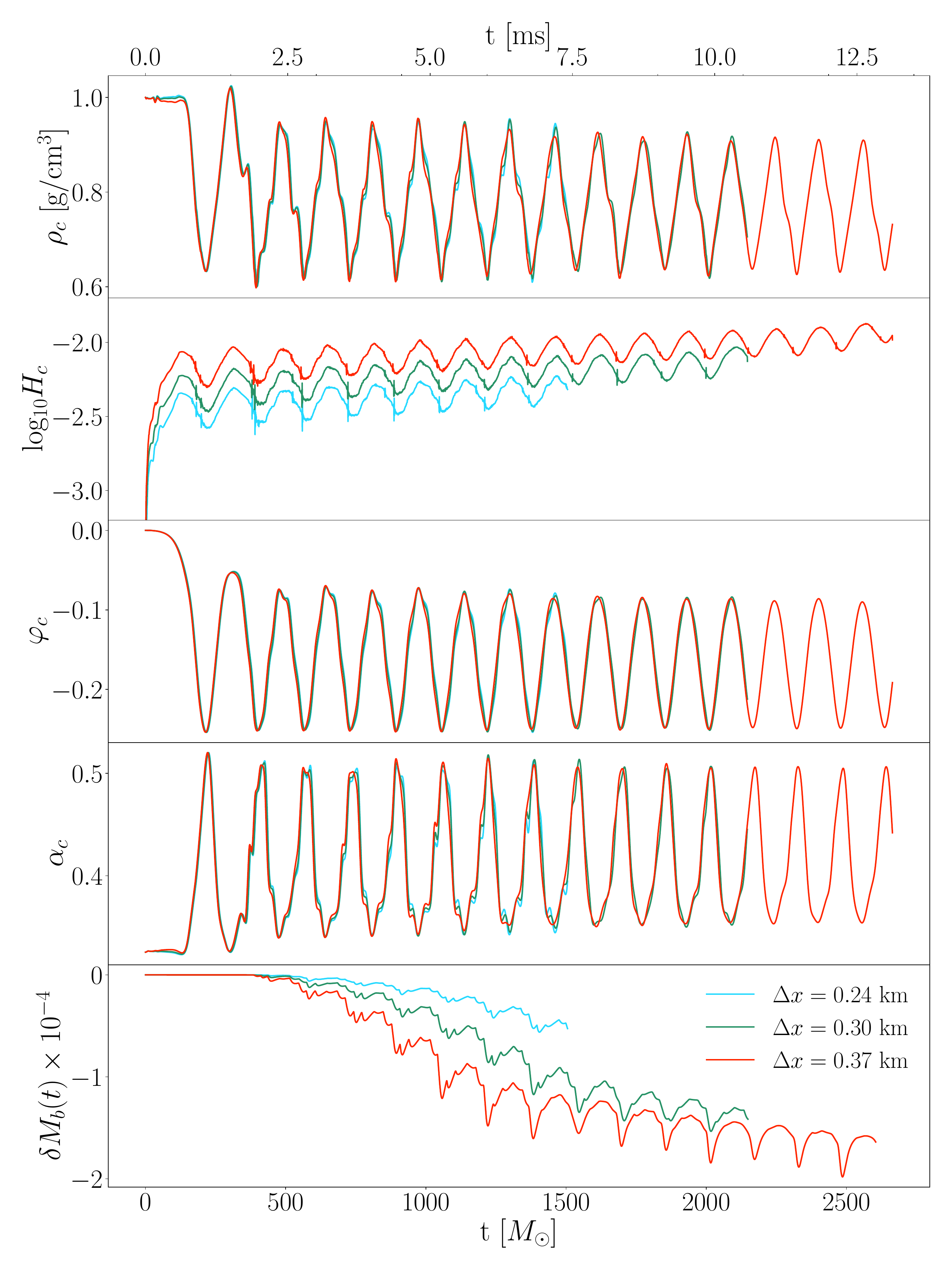}   
\caption{A star undergoing spontaneous scalarization. The figure describes the behavior of the star at the center. The top panel shows the change in relative density, the second row panel shows the Hamiltonian constraint, the third panel shows the scalar field, the fourth row panel shows the lapse function and the last panel displays the relative error for the baryonic mass. }
\label{fig:convergence_scalarizarion}
\end{figure}

\section{3+1 Decomposition in Cartesian coordinates}

We employ a $3+1$ decomposition of the spacetime metric to be able to express it in the following form:
\begin{equation}
ds^2 = -\alpha^2 dt^2 + \gamma_{ij}\left(dx^i + \beta^i dt\right)
                               \left(dx^j + \beta^j dt\right),
\end{equation}
where $\alpha$ is the lapse function, $\beta^i$ the shift vector, and
$\gamma_{ij}$ the spatial metric.

The lapse function is given by directly comparing with eq. \ref{eq:RNS_metric},
\begin{equation}
\alpha = e^{\frac{1}{2}(\xi+\sigma)},
\end{equation}
while the shift vector corresponds to a rigid rotation around the $z$--axis and has Cartesian components
\begin{equation}
\beta^x = \omega\, y, \qquad
\beta^y = -\omega\, x, \qquad
\beta^z = 0 ,
\end{equation}
where $\omega$ is the frame-dragging potential shown in eq. \ref{eq:RNS_metric}.

The spatial metric $\gamma_{ij}$ takes the form
\begin{align}
\gamma_{xx} &=
\frac{ e^{\xi-\sigma}\,y^2 + e^{2\delta} x^2}
     {x^2 + y^2}, \\[1ex]
\gamma_{yy} &=
\frac{e^{\xi-\sigma}\,x^2 + e^{2\delta} y^2}
     {x^2 + y^2}, \\[1ex]
\gamma_{xy} &=
\left( e^{2\delta} - e^{\xi-\sigma} \right)
\frac{x y}{x^2 + y^2}, \\[1ex]
\gamma_{xz} &= 0, \qquad \gamma_{yz} = 0, \\[1ex]
\gamma_{zz} &= e^{2\delta}.
\end{align}

On the rotation axis $(x=y=0)$ the metric is regular and reduces to the
conformally flat form
\begin{equation}
\gamma_{ij} = e^{2\delta}\,\delta_{ij}.
\end{equation}

Once the metric is computed, the extrinsic curvature tensor can be calculated by (eq. 1.57 in  ref. \cite{rot_stars_Gourgoulhon2010})
\begin{equation}
    \begin{aligned}
        K_{ij}=\frac{1}{2\alpha}\left( -\frac{\partial\gamma_{ij}}{\partial t} + \beta^k\frac{\partial\gamma_{ij}}{\partial x^k} +\gamma_{kj}\frac{\partial \beta^k}{\partial x^i} \gamma_{ki}\frac{\partial \beta^k}{\partial x^j}\right).
    \end{aligned}
\end{equation}

The scalar field is computed in the Einstein frame and later transformed as $\varphi=\sqrt{-2\beta_0} \,\phi$. The initial data for the conjugate momentum of the scalar field is calculated from Eq.~(\ref{eq:SF3+1}), assuming that $\partial_t\varphi=0$.

\section{Black hole diagnostics}

The most direct way of calculating the mass of a black hole is to look at their respective geometry since the size of the black hole is directly related to its mass: 

\begin{equation}{\label{eq:MassCeq}}
    M_{BH}=\frac{1}{4\pi}\int_0^{2\pi}d\phi\sqrt{g_{\phi\phi}} = \frac{C_{eq}}{4\pi}
\end{equation}

In order to estimate the ADM mass enclosed over a spherical surface, we calculate the following integral over spherical shells located at finite radii:

\begin{equation}{\label{eq:ADMmass}}
\begin{aligned}
        M_{ADM} = \oint_S dS^k  \alpha\gamma^{ij}\left( \partial_i\gamma_{jk}-\partial_k\gamma_{ij} \right) 
\end{aligned}
            \end{equation}

The total mass can also be found using the Christodoulou formula for a Kerr black hole \cite{Christodolou}:
\begin{equation}{\label{eq:Christodoulou}}
    M_{BH}^2=M_{irr}^2+\frac{4\pi J^2}{A}
\end{equation}

This approximation assumes that the spacetime is stationary and axisymmetric, so there is a degree of error before the black hole settles into some quasi-stationary state.

For the angular momentum calculation, we measure the horizon distortion by looking at the ratio between the polar and equatorial proper circunferences $C_p$ and $C_e$ of the horizon through the relation shown in ref. \cite{3dcollapse1}: 

\begin{equation}{\label{eq:J_fit}}
    \frac{J}{M^2} = \sqrt{1-\left(2.55\frac{C_p}{C_e}-1.55\right)^2}
\end{equation}

\end{document}